\documentclass[a4paper,11pt]{article}
\pdfoutput=1
\usepackage{jheppub}

\usepackage{amssymb,amsmath,amsfonts}
\usepackage[normalem]{ulem}
\usepackage[utf8x]{inputenc}
\usepackage{slashed}
\usepackage{graphicx}
\usepackage{tabularx}
\usepackage{here}
\usepackage{color}
\usepackage{csquotes} %fix backwards quotes
\usepackage{comment}
\usepackage{mathrsfs}
\usepackage{float}
\usepackage{ascmac}
\usepackage{multirow}
\usepackage{longtable}
\usepackage{bm}
\usepackage{ulem}

\usepackage[italicdiff]{physics}
\def\slash#1{\not\!\!#1}

\makeatletter
\newcommand*\rel@kern[1]{\kern#1\dimexpr\macc@kerna}
\newcommand*\widebar[1]{%
  \begingroup
  \def\mathaccent##1##2{%
    \rel@kern{0.8}%
    \overline{\rel@kern{-0.8}\macc@nucleus\rel@kern{0.2}}%
    \rel@kern{-0.2}%
  }%
  \macc@depth\@ne
  \let\math@bgroup\@empty \let\math@egroup\macc@set@skewchar
  \mathsurround\z@ \frozen@everymath{\mathgroup\macc@group\relax}%
  \macc@set@skewchar\relax
  \let\mathaccentV\macc@nested@a
  \macc@nested@a\relax111{#1}%
  \endgroup
}
\makeatother

\numberwithin{equation}{section}

\preprint{
\begin{minipage}{5cm}
\small
\flushright
EPHOU-26-13\\ 
\end{minipage}}

\title{Yukawa textures due to selection rules in magnetized orbifold models}

\author{Tatsuo Kobayashi,} 
\author{Tomoyasu Kori,} 
\author{Shuhei Miyamoto,} 
\author{Riku Nakano, \\ and} 
\author{Haruki Uchida}
\affiliation{
Department of Physics, Hokkaido University, Sapporo 060-0810, Japan\\
}

\emailAdd{kobayashi@particle.sci.hokudai.ac.jp}
\emailAdd{t-kori@particle.sci.hokudai.ac.jp}
\emailAdd{s-miyamoto@particle.sci.hokudai.ac.jp}
\emailAdd{r-nakano@particle.sci.hokudai.ac.jp}
\emailAdd{haru-uchida@particle.sci.hokudai.ac.jp}

\abstract{
We study systematically Yukawa textures, which can be derived from magnetized $T^2/\mathbf{Z}_2$ and $T^2/\mathbf{Z}_4$ orbifold models with various magnetic fluxes and Scherk-Schwarz phases. 
Remarkably, the nearest neighbor interaction pattern as well as the diagonal Yukawa matrix is statistically favorable in $T^2/\mathbf{Z}_2$ orbifold models. Also, in $T^2/\mathbf{Z}_4$ orbifold models, selection rules are nontrivial and lead to textures, which cannot be derived from group-theoretical symmetries.
}

\makeatletter
\gdef\@fpheader{}
\makeatother

\begin{document}

\maketitle

%%%%%%%%%%%%%%%%%%%%%%%%%%%%%%%%%%%%%%%%%%%%%%%%%%%%%%%%%%%%%%%%%
\section{Introduction}
\label{sec:Intro}

Symmetries are important in particle physics as well as other fields of physics.
In particular, symmetries determine which couplings among particles are allowed or forbidden in particle physics.
Conventionally, group-like symmetries have been used.

Indeed, string compactifications lead to selection rules due to group theory.
However, they also realize selection rules without group actions.
For example, in 
heterotic orbfold models, strings on the orbifold are specified by their boundary conditions, i.e.,
$X(\sigma=\pi)=gX(\sigma = 0)=\theta X(\sigma)+v$, 
where $g=(\theta,v)$ denote a space group element, and $\theta$ and $v$ denote a twist and a shift.
However, the same boundary condition can be satisfied by $hX$, i.e., $hX(\sigma=\pi)=ghX(\sigma = 0)$, where $h$ is another space group element.
That leads to $X(\sigma=\pi)=h^{-1}ghX(\sigma = 0)$.
That implies that string states are specified not by an element of space group, but its conjugacy class $h^{-1}gh$ \cite{Dixon:1986jc,Dixon:1986qv,Hamidi:1986vh}.
Multiplication rules are different between group elements and conjugacy classes.
Selection rules due to conjugacy class multiplication rules are different from those due to group theory and non-invertible.
See for concrete examples in heterotic orbifold models Refs.~\cite{Kobayashi:1990mc,Kobayashi:1991rp,Kobayashi:1995py,Kobayashi:2025ocp,Hernandez-Segura:2026yfx}.

Magnetized orbifold models also lead to non-invertible selection rules \cite{Kobayashi:2024yqq,Funakoshi:2024uvy}.
Compactifications with magnetic fluxes are interesting \cite{Bachas:1995ik,Blumenhagen:2000wh,Angelantonj:2000hi,Blumenhagen:2000ea,Cremades:2004wa}.
They lead to $M$ degenerate zero modes depending on magnetic fluxes.
Furthermore, magnetized $T^2$ models have the $\mathbf{Z}_M$ flavor symmetry \cite{Abe:2009vi,Berasaluce-Gonzalez:2012abm,Marchesano:2013ega}, and zero modes $\psi^j$ have $\mathbf{Z}_M$ charges $j=0,\cdots,M-1$.
Magnetized $T^2/\mathbf{Z}_2$ orbifold models are also interesting compactification models \cite{Abe:2008fi,Abe:2013bca,Abe:2014noa}.
On the $T^2/\mathbf{Z}_2$ orbifold, zero modes correspond to $\psi^j \pm \psi^{M-j}$ when Scherk-Schwarz (SS) phases are vanishing.
They behave as if they have both $\mathbf{Z}_2$ charges, $j$ and $-j$.
Thus, the $\mathbf{Z}_M$ is violated, but non-invertible selection rules remain.
Such non-invertible selection rules are called $\mathbf{Z}_2$ gauging of $\mathbf{Z}_M$ symmetries \cite{Kobayashi:2024cvp}, which also correspond to multiplication rules of 
the $\mathbf{Z}_2$ invariant part of $D_M\simeq \mathbf{Z}_M \rtimes \mathbf{Z}_2$ conjugacy classes. 
Other non-invertible selection rules are also studied.
(See e.g. Refs.~\cite{Kaidi:2024wio,Dong:2025jra}.)

Such non-invertible selection rules have several interesting applications in particle phenomenology.
First of all, they can realize interesting textures of mass matrices, which can not be derived simply by group theory
\cite{Kobayashi:2024cvp,Kobayashi:2025znw,Kobayashi:2025ldi,Jiang:2025psz,Qu:2026omn,Chen:2026mvi}.
Furthermore, specific textures are useful to solve the strong CP problem without axion \cite{Liang:2025dkm,Kobayashi:2025thd,Kobayashi:2025rpx}.
Also, non-invertible selection rules have been studied from various viewpoints such as neutrino physics, dark matter, multi-Higgs models, and standard model effective theory (SMEFT) \cite{Kobayashi:2025lar,Kobayashi:2025wty,Nomura:2026hcu,Suzuki:2025bxg,Suzuki:2025kxz,Suzuki:2025oov,Xu:2026nwh,Kobayashi:2025cwx,Nomura:2025sod,Chen:2025awz,Okada:2025kfm,Jangid:2025krp,Nomura:2025tvz,Okada:2026bpp, Kawamura:2026zae,Hamada:2026sui,Kobayashi:2026msq,Ding:2026sfm,Ding:2026bhp,Nakai:2025thw}.

Yukawa texture approaches were started by Steve Weinberg \cite{Weinberg:1977hb}.
Various patterns of textures have been proposed \cite{Fritzsch:1977vd,Fritzsch:1979zq,Ramond:1993kv,Fritzsch:2002ga,Xing:2015sva,Frampton:2002yf,Kageyama:2002zw}.
Texture zero structures are interesting, but zero entries are just assumed  conventionally.
However, non-invertible selection rules can derive certain patterns of textures as mentioned above.
Recently such studies on textures were extended to SMEFT operators \cite{Kobayashi:2026msq}.

The purpose of this paper is to study coupling selection rules and systematically examine which patterns of textures are derived in magnetized $T^2/\mathbf{Z}_2$ and $T^2/\mathbf{Z}_4$ orbifold models.
In magnetized $T^2$ models, there appear  exact $\mathbf{Z}_M$ symmetries, which control Yukawa couplings and other couplings, as mentioned above.
In orbifold models, exact $\mathbf{Z}_M$ symmetries are violated, but some selection rules may remain.
Indeed, magnetized $T^2/\mathbf{Z}_2$ orbifold models without SS phases are typical models to lead to non-invertible selection rules, i.e., $\mathbf{Z}_2$ gauging of $\mathbf{Z}_M$ symmetries.
Thus, we consider magnetized $T^2/\mathbf{Z}_2$ orbifold models with all the possible SS phases, although non-invertible selection rules were studied in models with vanishing SS phases \cite{Kobayashi:2024yqq} so far.
We will show that models with SS phases have selection rules similar to models without SS phases.
Within such a framework, we study all the possible configurations of magnetic fluxes, SS phases, and $\mathbf{Z}_2$ parities to lead to three zero modes, i.e., three generations.
They are assigned to three generations of left-handed and right-handed fermions, where assignments of left-handed and right-handed are not always same.
Then, we investigate systematically possible mass textures by taking the possible Higgs modes into account.
We also study magnetized $T^2/\mathbf{Z}_4$ orbifold models.
The above selection rules in $T^2/\mathbf{Z}_2$ models such as $\mathbf{Z}_2$ gauging of $\mathbf{Z}_M$ symmetries would be further violated by the $\mathbf{Z}_4$ twist of 
the $T^2/\mathbf{Z}_4$ orbifold.
Any selection rules may not remain in magnetized $T^2/\mathbf{Z}_4$ orbifold models.
We examine 
whether some selection rules originating from $\mathbf{Z}_M$ symmetries remain in magnetized $T^2/\mathbf{Z}_4$ orbifold models with and without SS phases, by investigating the possible textures in magnetized $T^2/\mathbf{Z}_4$ orbifold models.

The rest of this paper is organized as follows.
In section 2, we review magnetized $T^2$ models and $T^2/\mathbf{Z}_2$ orbifold models.
In section 3, we study their selection rules with non-vanishing SS phases, and systematically examine the possible textures in magnetized $T^2/\mathbf{Z}_2$ orbifold models.
In section 4, we study coupling selection rules and possible textures in magnetized $T^2/\mathbf{Z}_4$ orbifold models with and without SS phases.
In section 5, we also study additional cancellations in Yukawa matrices by explicit computations.
Section 6 is devoted to conclusion.

%%%%%%%%%%%%%%%%%%%%%%%%%%%%%%%%%%%%%%%%%%%%%%%%%%%%%%%%%%%%%%%%%
\section{Magnetized $T^2/\mathbf{Z}_2$ orbifold models}
\label{sec:orbifold}

Here, we review briefly magnetized $T^2$ and  $T^2/\mathbf{Z}_2$ orbifold models.
In $T^2$ models, the exact $\mathbf{Z}_M$ symmetry appears.
On the $T^2/\mathbf{Z}_2$ orbifold, the $\mathbf{Z}_M$ symmetry is violated, but some selection rules remain.
In Ref.~\cite{Kobayashi:2024yqq,Kobayashi:2024cvp}, it was shown that $\mathbf{Z}_2$ gauging of $\mathbf{Z}_M$ symmetries can be realized in magnetized $T^2/\mathbf{Z}_2$ models without SS phases.

\subsection{Magnetized $T^2$ models}

We consider higher dimensional gauge theory, where magnetic fluxes along $U(1)$ directions of the gauge group $G$ are induced in extra compact dimensions.
The simple compactification is $T^2$.
For simplicity, we start with $U(1)$ gauge theory on $T^2$ with magnetic fluxes. 

A two-dimensional torus $T^2$ is constructed by identifying the complex plane $\mathbf{C}$ by a lattice $\Lambda$,  $T^2\simeq\mathbf{C}/\Lambda$. Two basis of $\Lambda$ are $e_1=1$ and $e_2=\tau$, where $\tau\in\mathbf{C}$ denotes the complex structure modulus. We use the complex
coordinate $z=x+\tau y\in\mathbf{C}$ where $x$ and $y$ are real coordinates. Since we start with $U(1)$ gauge theory, we can introduce an Abelian magnetic flux such that $\int_{T^2} F=2\pi M$;
\begin{align}
    F=\dfrac{\pi i M}{\mathrm{Im}\tau}dz\wedge d\overline{z},
\end{align}
where Dirac’s quantization condition requires $M\in\mathbf{Z}$. 

\subsubsection{Zero modes and coupling selection rules}

For a two-dimensional spinor $(\psi_{+},\psi_{-})^{\mathrm{T}}$ in $T^2$ transforming under $U(1)$ gauge field with charge $q=1$, we can derive the zero modes of the Dirac equation $i\slash{D}\psi=0$. We introduce SS phase $(\alpha_{1},\alpha_{\tau})$ satisfying the following boundary conditions:
\begin{align}
\psi^{j,M}_{\pm,(\alpha_1,\alpha_{\tau})}(z+ 1,\tau) &= e^{2\pi i\alpha_1}e^{\pi iM\frac{\mathrm{Im}z}{\mathrm{Im}\tau}}\psi^{j,M}_{\pm,(\alpha_1,\alpha_{\tau})}(z,\tau),\\
\psi^{j,M}_{\pm,(\alpha_1,\alpha_{\tau})}(z+ \tau,\tau) &= e^{2\pi i\alpha_2}e^{\pi iM\frac{\mathrm{Im}(\bar{\tau}z)}{\mathrm{Im}\tau}}\psi^{j,M}_{\pm,(\alpha_1,\alpha_{\tau})}(z,\tau).
\end{align}

The Dirac equation has $|M|$ independent solutions, each of them having a different wavefunction $\psi^{j,|M|}_{\pm,(\alpha_1,\alpha_{\tau})}$, where $j=0,1,\cdots, |M|-1.$ 
If  $M>0,$ $\psi^{j,M}_{+,(\alpha_1,\alpha_{\tau})}$ has the $M$ degenerate solutions.
They are written as \cite{Cremades:2004wa}
\begin{align}
    \psi^{j,M}_{+,(\alpha_1,\alpha_{\tau})}(z,\tau)=\mathcal{N}\exp\left(
    \pi i M \frac{z\,\mathrm{Im}\,z}{\mathrm{Im}\,\tau}\right)\vartheta
    \begin{bmatrix}
        \dfrac{j+\alpha_1}{M} \\
        -\alpha_{\tau}
    \end{bmatrix}
    (Mz,M\tau).
    \label{eq:wavefunction}
\end{align}
Here, $\mathcal{N}$ is the normalization factor. It can be determined by imposing the following normalization condition

\begin{align}
\int_{T^2} dz\, d\bar{z}\,
\left(\psi^{j,M}_{+,(\alpha_1,\alpha_{\tau})}\right)^{*}
\psi^{k,M}_{+,(\beta_1,\beta_{\tau})}
=
(2\,\mathrm{Im}\,\tau)^{-1/2}
\delta_{j,k},
\end{align}
where $\alpha_i \equiv \beta_i$ (mod 1) for $i=1,\tau$.
$\vartheta$ in eq.\eqref{eq:wavefunction} denotes the Jacobi theta function defined by 
\begin{align}
    \vartheta
    \begin{bmatrix}
        a\\
        b
    \end{bmatrix}
    (\nu,\tau)=\sum_{\ell\in\mathbf{Z}}
    e^{\pi i (a+\ell)^2\tau}
    e^{2\pi i (a+\ell)(\nu+b)}.
\end{align}
The wavefunctions for $M<0$ can be obtained in an analogous way. The solutions $\psi^{j,M}_{+,(\alpha_1,\alpha_{\tau})}$ and $\psi^{j,M}_{-,(\alpha_1,\alpha_{\tau})}$ are mutually exclusive, since the $\vartheta$-function defining $\psi^{j,M}_{+,(\alpha_1,\alpha_{\tau})}$ converges only when $M>0$, while $\psi^{j,M}_{-,(\alpha_1,\alpha_{\tau})}$ is well-defined only for $M<0$.
In what follows, we consider the models with $M>0$, where only $\psi^{j,M}_{+,(\alpha_1,\alpha_{\tau})}$ remains and we omit the chirality index $+$.

The bosonic modes have the same wave functions as the fermionic modes \cite{Cremades:2004wa}.
The Yukawa coupling constants are determined by the overlap integrals
of wavefunctions in the extra dimensions.
For example, for models with vanishing SS phases,
the three-point couplings on $T^2$
are calculated from the following expression:
\begin{align}
y_{i j \bar{a}}
=
g \int d^2 z \,
\psi^{\,i,M_1}(z)
\psi^{\,j,M_2}(z)
\left(
\psi^{\,a,M_3}(z)
\right)^{*},
\end{align}
up to normalization of the kinetic factors, 
where $g$ denotes the coupling in higher dimensions.
Here, we consider the case in which SS phases vanish. 
Gauge invariance requires the condition for magnetic fluxes
\begin{align}
M_1+M_2=M_3.
\end{align}
The result obtained by explicit calculation is \cite{Cremades:2004wa}
\begin{align}
y_{i j \bar{a}}
=g \sum_{m \in \mathbf{Z}_{M_3}}
\delta_{i+j+M_1 m,\; a}
\,
\vartheta
\begin{bmatrix}
\dfrac{M_2 i - M_1 j + M_1 M_2 m}
{M_1 M_2 M_3}
\\[6pt]
0
\end{bmatrix}
\!\left(
0,\;
\tau M_1 M_2 M_3
\right).
\end{align}

Note that the integer $m$ appearing in the Kronecker delta
is defined modulo $M_3$.
The Kronecker-delta factor yields the selection rule
for allowed couplings,
\begin{align}
i+j-a
=
M_3 l - M_1 m,
\quad
m\in\mathbf{Z}_{M_3},
\quad
l\in\mathbf{Z}_{M_1}.
\end{align}
When $\mathrm{GCD}(M_1,M_2,M_3)=M$,
this selection rule is simplified to
\begin{align}
i+j-a
\equiv
0
\pmod{M}.
\label{eq:selection}
\end{align}
This implies that  $\mathbf{Z}_M$ charges can be defined
for the quantum numbers $i,j,a$ of the zero modes,
and the allowed couplings are controlled by this
$\mathbf{Z}_M$ symmetry.
In fact, each quantum number $i,j,a$
corresponds to a quantized momentum
with a modulo structure defined by the magnetic flux $M_i$.
When $\mathrm{GCD}(M_1,M_2,M_3)=M$,
this modulo structure becomes $\mathbf{Z}_M$,
and the conservation law of these discrete momenta
corresponds to the condition required by
$\mathbf{Z}_M$ invariance.

Let $M$ denote the greatest common divisor of the background fluxes
$M_1,\cdots,M_L$.
These modes carry $\mathbf{Z}_M$ charges, and couplings among these
modes are allowed when $\mathbf{Z}_M$ invariance is satisfied.
For example, when $M_1=M$, the mode $\psi^{\,i,M}$
has $M$ zero modes $(i=0,\cdots,M-1)$.
In this case, the $\mathbf{Z}_M$ transformation of the zero modes is
given by
\begin{align}
\psi^{\,i,M}
\;\longrightarrow\;
Z\,\psi^{\,i,M},
\end{align}
where $Z$ is a representation of $\mathbf{Z}_M$ given by
\begin{align}
Z=\begin{pmatrix}
        1 & & & & \\
          & g & & \\
          &  & g^2 & \\
          &  &  &  \ddots \\
          &  &  &  & g^{M-1}
    \end{pmatrix},
\qquad
g=e^{2\pi i/M}.
\end{align}

This is a cyclic group of order $M$,
which directly represents the conservation law of discrete momenta.

\subsubsection{Yukawa texture}

The three generations can be realized only if $M=3$.
They have $\mathbf{Z}_3$ charges, 0,1,2, as mentioned in the previous section.
That is, the $\mathbf{Z}_3$ symmetry determines allowed couplings.
Suppose that both of left-handed and right-handed fermions have $\mathbf{Z}_3$ charges, 0,1,2.
Then, the $\mathbf{Z}_3$ symmetry leads to the following textures:
\begin{align}\label{eq:Y-T2-1}
    \begin{pmatrix}
        * & 0 & 0 \\
        0 & 0 & * \\
        0 & * & 0
    \end{pmatrix},
\end{align}
when the Higgs modes has $\mathbf{Z}_3$ charge 0,
\begin{align}\label{eq:Y-T2-2}
    \begin{pmatrix}
        0 & 0 & * \\
        0 & * & 0 \\
        * & 0 & 0
    \end{pmatrix},
\end{align}
when the Higgs modes has $\mathbf{Z}_3$ charge 1,
\begin{align}\label{eq:Y-T2-3}
    \begin{pmatrix}
        0 & * & 0 \\
        * & 0 & 0 \\
        0 & 0 & *
    \end{pmatrix},
\end{align}
when the Higgs modes has $\mathbf{Z}_3$ charge 2.
If the Higgs field is one of them, one can not realize mixing angles, and 
these textures are not realistic.
For example, we may need to combine them by assuming multi-Higgs models.

\subsubsection{$T^2$ models with SS phases}

Similarly, we can study models with non-vanishing SS phases.
The Yukawa couplings among zero modes with SS phases are written by \cite{Cremades:2004wa,Funakoshi:2024uvy}
\begin{align}
y^{i,j,a}_{T^2}&=\int_{T^2} dzd\bar{z}\;
\psi^{(i+\alpha^{(i)}_1,\alpha^{(i)}_\tau),M_i}_{T^2}(z)\,
\psi^{(j+\alpha^{(j)}_1,\alpha^{(j)}_\tau),M_j}_{T^2}(z)\,
\left(\psi^{(a+\alpha^{(a)}_1,\alpha^{(a)}_\tau),M_a}_{T^2}(z)
\right)^*\nonumber\\
&=c(M_i,M_j,M_a)
\exp\left[
2\pi i\left(
\frac{(i+\alpha^{(i)}_1)\alpha^{(i)}_\tau}{M_i}
+\frac{(j+\alpha^{(j)}_1)\alpha^{(j)}_\tau}{M_j}
-\frac{(a+\alpha^{(a)}_1)\alpha^{(a)}_\tau}{M_a}
\right)
\right]
\nonumber\\
&\quad\times
\sum_{m=0}^{M_a-1}
\vartheta
\begin{bmatrix}
\dfrac{M_j(i+\alpha^{(i)}_1)-M_i(j+\alpha^{(j)}_1)+M_iM_jm
}{M_iM_jM_a}\\0
\end{bmatrix}
\left(M_j\alpha^{(i)}_\tau-M_i\alpha^{(j)}_\tau,\,
M_iM_jM_a\tau\right)
\nonumber\\&\quad\times\delta_{(i+\alpha^{(i)}_1)+(j+\alpha^{(j)}_1)-(a+\alpha^{(a)}_1),\,M_a\ell-M_im}.
\label{eq:yukawa-coupling}
\end{align}
Hence, the coupling selection rules are written by
\begin{eqnarray}
 (i+\alpha_1^{(i)})+ (j+\alpha_1^{(j)})- (a+\alpha_1^{(a)})=0\quad(\mathrm{mod}\ M),
\label{eq:selection-T2}
\end{eqnarray}
where $M$ is the greatest common divisor of $M_a$ and $M_i$. 
In $T^2/\mathbf{Z}_2$ orbifold models, the SS phases are constrained to be $\alpha_1=0, 1/2$, as will be mentioned in the next subsection. 
For $\alpha_1=0, 1/2$, the $\mathbf{Z}_{2M}$ symmetry appears.

By introducing continuous SS phases, the number of zero modes does not change.
That is, we need $M=3$ to realize the three generation models.
In addition, they lead the same textures as Eqs.~(\ref{eq:Y-T2-1}),(\ref{eq:Y-T2-2}), (\ref{eq:Y-T2-3}).

\subsection{Magnetized $T^2/\mathbf{Z}_2$ orbifold models}

Let us discuss the orbifold $T^2/\mathbf{Z}_2$ by imposing the identification $z\sim -z$ on $T^2$. In this case, the possible combinations of SS phases $(\alpha_{1},\alpha_{\tau})$ are the following four patterns:
\begin{align}
    (0,0),\quad (0,1/2),\quad (1/2,0), \quad (1/2,1/2).
\end{align}

By taking linear combinations of the wavefunctions on $T^2$, one obtains the $\mathbf{Z}_2$ even and odd modes on $T^2/\mathbf{Z}_2$ \cite{Abe:2013bca,Abe:2008fi}, 
\begin{align}
    \psi^{(j+\alpha_1^{(j)}, \alpha_\tau^{(j)}), M}_{+} (z) = \psi^{(j+\alpha_1^{(j)}, \alpha_\tau^{(j)}), M}_{T^2} (z) + e^{-4\pi i (j+\alpha_1^{(j)}) \alpha_\tau^{(j)} / M} \psi^{(M-(j+\alpha_1^{(j)}), \alpha_\tau^{(j)}), M}_{T^2} (z),
\end{align}
\begin{align}
    \psi^{(j+\alpha_1^{(j)}, \alpha_\tau^{(j)}), M}_{-} (z) = \psi^{(j+\alpha_1^{(j)}, \alpha_\tau^{(j)}), M}_{T^2} (z) - e^{-4\pi i (j+\alpha_1^{(j)}) \alpha_\tau^{(j)} / M} \psi^{(M-(j+\alpha_1^{(j)}), \alpha_\tau^{(j)}), M}_{T^2} (z).    
\end{align}
Here, the normalization factors are omitted. 

Rewriting the three-point couplings of the $\mathbf{Z}_2$ even modes for the case of magnetic fluxes $M$-$M$-$2M$ in terms of the three-point couplings on $T^2$, one obtains

\begin{align}
    y_{T^2/\mathbf{Z}_2}^{i,j,a} &= \int_{T^2/\mathbf{Z}_2} dz d\bar{z} \psi^{(i+\alpha_1^{(i)}, \alpha_\tau^{(i)}), M}_{+} \cdot \psi^{(j+\alpha_1^{(j)}, \alpha_\tau^{(j)}), M}_{+} \cdot \left( \psi^{(a+\alpha_1^{(a)}, \alpha_\tau^{(a)}), 2M}_{+} \right)^*\nonumber\\
    &= y_{T^2}^{i,j,a} + (u^{(a)})^{1/2} y_{T^2}^{i,j, 2M-a} + u^{(j)} y_{T^2}^{i, M-j, a} + u^{(j)}(u^{(a)})^{1/2} y_{T^2}^{i, M-j, 2M-a} + u^{(i)} y_{T^2}^{M-i, j, a} \nonumber\\&+ u^{(i)}(u^{(a)})^{1/2} y_{T^2}^{M-i, j, 2M-a}+ u^{(i)}u^{(j)} y_{T^2}^{M-i, M-j, a} + u^{(i)}u^{(j)}(u^{(a)})^{1/2} y_{T^2}^{M-i, M-j, 2M-a},
\end{align}
where,
\begin{eqnarray}
&    u^{(i)} = e^{-4\pi i (i+\alpha_1^{(i)}) \alpha_\tau^{(i)} / M}, \quad u^{(j)} = e^{-4\pi i (j+\alpha_1^{(j)}) \alpha_\tau^{(j)} / M}, \notag \\ & u^{(a)} = e^{-4\pi i (a+\alpha_1^{(a)}) \alpha_\tau^{(a)} / (2M)}.
\end{eqnarray}
Therefore, the selection rules for non-vanishing three-point coupling $y_{T^2/\mathbf{Z}_2}^{i,j,a}$ on $T^2/\mathbf{Z}_2$ with SS phase is given by
\begin{eqnarray}
\pm (i+\alpha_1^{(i)})\pm (j+\alpha_1^{(j)})\pm (a+\alpha_1^{(a)})=0\quad(\mathrm{mod}\ M).
\label{eq:selection}
\end{eqnarray}
Comparing this with the selection rule \eqref{eq:selection-T2} for the three-point couplings on $T^2$, the coefficients in front of each charge become $\pm$, leading to a slightly weaker condition: it is sufficient to choose one of the signs appropriately so that the sum of charges vanishes.
Similarly, we can study three-point couplings including $\mathbf{Z}_2$ odd modes. 
Indeed, only the three-point couplings among two $\mathbf{Z}_2$ odd modes and a $\mathbf{Z}_2$ even modes are allowed.
They also satisfy the same selection rule (\ref{eq:selection}).

Our purpose is to study systematically which patterns of Yukawa textures are derived in magnetized $T^2/\mathbf{Z}_2$ orbifold models.
For such purpose, we classify combinations of left-handed and right-handed fermions and Higgs modes.
All of the possible combinations of $\mathbf{Z}_2$-parity with the combinations of SS phase in terms of overlap integral on $T^2/\mathbf{Z}_2$ are 15 patterns of Tables, \ref{tab1}-\ref{tab5}, where  $(\alpha_1^{(i)},\alpha_{\tau}^{(i)})+(\alpha_1^{(j)},\alpha_{\tau}^{(j)})=(\alpha_1^{(a)},\alpha_{\tau}^{(a)})$\ $(\mathrm{mod}\ 1)$.
\begin{table}[H]
    \centering
    \begin{tabular}{|c|c|c|c|} \hline
        SS phase & $(0,0)$ & $(0,0)$ & $(0,0)$ \\\hline
       Pattern 1 & even & even & even  \\\hline
       Pattern 2 & even & odd & odd \\\hline
    \end{tabular}
    \caption{The possible combinations of $\mathbf{Z}_2$-parity with the combinations of SS phase $((0,0),\ (0,0),\ (0,0))$.}
    \label{tab1}
\end{table}

\begin{table}[H]
    \centering
    \begin{tabular}{|c|c|c|c|} \hline
        SS phase & $(0,0)$ & $(1/2,0)$ & $(1/2,0)$ \\\hline
        Pattern 3 & even & even & even \\\hline
        Pattern 4 & even & odd & odd  \\\hline
        Pattern 5 & odd & even & odd \\\hline
    \end{tabular}
    \caption{The possible combinations of $\mathbf{Z}_2$-parity with the combinations of SS phase $((0,0),\ (1/2,0),\ (1/2,0))$.}
    \label{tab2}
\end{table}

\begin{table}[H]
    \centering
    \begin{tabular}{|c|c|c|c|} \hline
        SS phase & $(0,0)$ & $(0,1/2)$ & $(0, 1/2)$ \\\hline
        Pattern 6 & even & even & even \\\hline
        Pattern 7 & even & odd & odd \\\hline
        Pattern 8 & odd & even & odd \\\hline
    \end{tabular}
    \caption{The possible combinations of $\mathbf{Z}_2$-parity with the combinations of SS phase $((0,0),\ (0,1/2),\ (0,1/2))$.}
    \label{tab3}
\end{table}

\begin{table}[H]
    \centering
    \begin{tabular}{|c|c|c|c|} \hline
        SS phase & $(0,0)$ & $(1/2,1/2)$ & $(1/2,1/2)$ \\\hline
        Pattern 9 & even & even & even \\\hline
        Pattern 10 & even & odd & odd \\\hline
        Pattern 11& odd & even & odd \\\hline
    \end{tabular}
    \caption{The possible combinations of $\mathbf{Z}_2$-parity with the combinations of SS phase $((0,0),\ (1/2,1/2),\ (1/2,1/2))$.}
    \label{tab4}
\end{table}

\begin{table}[H]
    \centering
    \begin{tabular}{|c|c|c|c|} \hline
        SS phase & $(0,1/2)$ & $(1/2,0)$ & $(1/2,1/2)$ \\\hline
        Pattern 12 & even & even & even \\\hline
        Pattern 13 & even & odd & odd \\\hline
        Pattern 14& odd & even & odd \\\hline
        Pattern 15& odd & odd & even \\\hline
    \end{tabular}
    \caption{The possible combinations of $\mathbf{Z}_2$-parity with the combinations of SS phase $((0,1/2),\ (1/2,0),\ (1/2,1/2))$.}
    \label{tab5}
\end{table}

The values of the magnetic fluxes $M$ that reproduce the three-generation structure for each combination of SS phase and $\mathbf{Z}_2$-parity are listed in Table \ref{tab:three_gen}.
The generation number depends on the flux $M$, SS phases as well as $\mathbf{Z}_2$ parities in orbifold models, although it depends only on the flux $M$, but not on SS phase in $T^2$ models.
The selection rules also depend on the flux $M$, SS phases, and $\mathbf{Z}_2$ parities in $T^2/\mathbf{Z}_2$ orbifold models.
Thus, we study systematically Yukawa textures in magnetized $T^2/\mathbf{Z}_2$ models in the next section, although the limited models with $M=4,5$, vanishing SS phases, and $\mathbf{Z}_2$ even parity have been studied in the literature.

\begin{table}[h]
\centering
\begin{tabular}{|c|c|c|}
\hline
 SS phase & $\mathbf{Z}_2$-parity & Magnetic fluxes $M$ \\ \hline
\multirow{2}{*}{$(0,0)$} & even & 4,\ 5 \\ \cline{2-3}
                   & odd & 7,\ 8 \\ \hline

\multirow{2}{*}{$(1/2,0)$} & even & 5,\ 6 \\ \cline{2-3}
                   & odd  & 6,\ 7 \\ \hline

\multirow{2}{*}{$(0,1/2)$} & even & 5,\ 6 \\ \cline{2-3}
                   & odd & 6,\ 7 \\ \hline

\multirow{2}{*}{$(1/2,1/2)$} & even & 6,\ 7 \\ \cline{2-3}
                   & odd & 5,\ 6 \\ \hline
\end{tabular}
\caption{SS phases, $\mathbf{Z}_2$-parities, and magnetic fluxes $M$
yielding a three-generation structure.}
\label{tab:three_gen}
\end{table}

%%%%%%%%%%%%%%%%%%%%%%%%%%%%%%%%%%%%%%%%%%%%%%%%%%%%%%%%%%%%%%%%%
\section{Selection rules and Yukawa textures in magnetized $T^2/\mathbf{Z}_2$ orbifold models}
\label{sec:SI-texture}

Here, we study selection rules and Yukawa textures derived from magnetized $T^2/\mathbf{Z}_2$ orbifold models.

\subsection{Non-invertible multiplication rules}
\label{sec:NIMRs}

As reviewed in the previous section, the selection rules in magnetized $T^2/\mathbf{Z}_2$ orbifold models are different from those in magnetized $T^2$ models.
When SS phases vanish, the selection rules on magnetized $T^2$ models correspond to the $\mathbf{Z}_M$ symmetry, where $\psi^{j,M}$ has the $\mathbf{Z}_M$ charge $j$ and transforms as $\psi^{i,M} \to e^{2\pi j/M}\psi^{j,M}$.
On the other hand, the zero modes on $T^2/\mathbf{Z}_2$ orbifold are written by 
$\psi^{j,M}_\pm = \psi^{j,M} \pm \psi^{M-j,M}$.
The $\mathbf{Z}_M$ symmetry is violated.
The zero modes $\psi^{j,M}_\pm$  behave as if they have both $\mathbf{Z}_M$ charges, $j$ and $M-j$.
That leads to the selection rules (\ref{eq:selection}).
These selection rules can be represented as follows.
We define the  following class:
\begin{align}
    [g^k]=\{hg^kh^{-1}|h=e,r\}=\{g^k, g^{-k}\},
\end{align}
where $g$ denotes the generator of $\mathbf{Z}_M$ symmetry, and $r$ is the element of $\mathbf{Z}_2$ symmetry. It corresponds to geometrical $\mathbf{Z}_2$ orbifolding. 
That is called $\mathbf{Z}_2$ gauging of $\mathbf{Z}_M$ symmetries in Ref.~\cite{Kobayashi:2024cvp} and that also corresponds to $\mathbf{Z}_2$ invariant conjugacy classes of $D_M \simeq \mathbf{Z}_M \rtimes \mathbf{Z}_2$.
These classes correspond to the zero modes on the $T^2/\mathbf{Z}_2$ orbifold.
The multiplication rules of these classes are obtained as 
\begin{align}
    &[g^{k}][g^{k'}]=[g^{k+k'}]+[g^{k-k'}].
\end{align}
That provides us with the selection rules of zero modes on the $T^2/\mathbf{Z}_2$ orbifold.

The above discussion on modes without SS phases can be extended to modes with SS phase.
Since the case with SS phases include a charge $k+\alpha^{(k)}_1$ from $\psi_{+}^{(k+\alpha^{(k)}_1, \alpha^{(k)}_{\tau})}$, we also consider the following classes with half charges:
\begin{align}
    [g^{k+1/2}]=\{g^{k+1/2},g^{-k-1/2}\}.
\end{align}
These classes satisfy the following non-invertible multiplication
rules:
\begin{align}
%    &[g^{k}][g^{k'}]=[g^{k+k'}]+[g^{k-k'}]\\
    &[g^{k+1/2}][g^{k'}]=[g^{k+k'+1/2}]+[g^{k-k'+1/2}]\\
    &[g^{k+1/2}][g^{k'+1/2}]=[g^{k+k'+1}]+[g^{k-k'}].
\end{align}

For example, when $M=5$ for patterns 3 in Table \ref{tab2}, there are three classes $[g^0],[g^1],[g^2]$ for zero-modes wavefunctions $\psi_{+}^{(k+0,0)}$, and three classes $[g^{1/2}],[g^{3/2}],[g^{5/2}]$ for $\psi_{+}^{(k'+1/2,0)}$ and $\psi_{+}^{(k''+1/2,0)}$. They satisfy the following fusion rules:

\begin{align*}
  &[g^{0}] [g^{0}] = [g^{0}], \quad 
  [g^{0}][g^{1}] = [g^{1}], \quad
  [g^{0}] [g^{2}] = [g^{2}], \\
  &[g^{1}] [g^{1}] = [g^{0}] + [g^{2}], \quad
  [g^{1}] [g^{2}] = [g^{1}] + [g^{2}], \quad
  [g^{2}] [g^{2}] = [g^{0}] + [g^{1}],
\end{align*}
for $[g^k][g^{k'}]$ with integer $k$ and $k'$,
\begin{align*}
  &[g^{0}] [g^{1/2}] = [g^{1/2}], \quad
  [g^{0}] [g^{3/2}] = [g^{3/2}], \quad 
  [g^{0}] [g^{5/2}] = [g^{5/2}] \\
  &[g^{1}] [g^{1/2}] = [g^{1/2}] + [g^{3/2}], \quad 
  [g^{1}] [g^{3/2}] = [g^{1/2}] + [g^{5/2}], \quad 
  [g^{1}] [g^{5/2}] = [g^{3/2}] \\
  &[g^{2}] [g^{1/2}] = [g^{3/2}] + [g^{5/2}], \quad
  [g^{2}] [g^{3/2}] = [g^{1/2}] + [g^{3/2}], \quad
  [g^{2}] [g^{5/2}] = [g^{1/2}],
\end{align*}
for $[g^k][g^{k'+1/2}]$ with integer $k$ and $k'$,
\begin{align*}
  &[g^{1/2}] [g^{1/2}] = [g^{0}] + [g^{1}], \quad 
  [g^{1/2}] [g^{3/2}] = [g^{1}] + [g^{2}], \quad
  [g^{1/2}] [g^{5/2}] = [g^{2}] \\
  &[g^{3/2}] [g^{1/2}] = [g^{1}] + [g^{2}], \quad 
  [g^{3/2}] [g^{3/2}] = [g^{0}] + [g^{2}], \quad 
  [g^{3/2}] [g^{5/2}] = [g^{1}] \\
  &[g^{5/2}] [g^{1/2}] = [g^{2}], \quad 
  [g^{5/2}] [g^{3/2}] = [g^{1}], \quad 
  [g^{5/2}] [g^{5/2}] = [g^{0}]
\end{align*}
for $[g^{k+1/2}][g^{k'+1/2}]$ with integer $k$ and $k'$,
From the viewpoint of conjugacy classes, 
selection rules \eqref{eq:selection} imply that the couplings are allowed if the trivial conjugacy class $[g^0]$ appears on the right-hand side of the non-invertible multiplications of conjugacy classes.

%%%%%%%%%%%%%%%%%%%%%%%%%%%%%%%%%%%%%%%%%%%%%%%%%%%%%%%%%%%%%%%%%
\subsection{Yukawa textures} 
\label{sec:Textures}

Here, we study all the possible textures derived in magnetized $T^2/\mathbf{Z}_2$ orbifold models.
Three zero modes are realized only by eight combinations of SS phase, $\mathbf{Z}_2$ parities, and magnetic fluxes shown in Table \ref{tab:three_gen}.
Left handed and right handed fermions can be assigned to one of eight combinations. 
There are totally 64 combinations.
After we fix one of these 64 combinations of left and right handed fermions, we chose the Higgs field such that their Yukawa couplings are allowed as shown in Tables \ref{tab1}-\ref{tab5}.

We show some examples.
We consider the magnetic flux $M=5$.
Then, we study the models, where both left-handed and right-handed fermions are $\mathbf{Z}_2$ even modes and have vanishing SS phases.
The three generations of both left-handed and right-handed fermions correspond to the following three classes:
\begin{align}
\label{eq:flavor-assign-M=3-1}
    \text{i} : ([g^0],[g^1],[g^2]). 
\end{align}
In this case, the Higgs mode must also correspond to 
one of $[g^0]$, $[g^1]$, and $[g^2]$ to obtain allowed Yukawa couplings.
Yukawa textures are shown in Table \ref{tab:Y-M=5-1}, where the asterisk symbols "$*$" denote non-vanishing elements.
The Yukawa texture is diagonal when the Higgs mode corresponds to $[g^0]$.
This texture can be derived by group-like symmetry when we assign $\mathbf{Z}_N$ charges or $U(1)$ charges to three generations of left-handed and right-handed fermions as well as the Higgs mode.
On the other hand, the nearest neighbor interaction (NNI) pattern \cite{Branco:1988iq,Branco:1994jx,Branco:1999nb} can be realized when the Higgs mode 
corresponds to $[g^1]$.
This texture can not be derived by group-like symmetry  unless the symmetry is broken and/or multi-Higgs modes are introduced.
The third texture, which is obtained when the Higgs mode corresponds to $[g^2]$, is just a permutation of  rows and columns from the NNI pattern.

\begin{table}[H]
    \centering
    \begin{tabular}{|c|c|c|c|} \hline
      Flavor (left,right)  & Higgs $[g^{0}]$ & Higgs $[g^{1}]$ & Higgs $[g^{2}]$\\ \hline
\begin{tabular}{l}
(i,i) 
\end{tabular} 
& 
$\begin{pmatrix}
* & 0 & 0 \\
0 & * & 0 \\
0 & 0 & *
\end{pmatrix}$
& 
$\begin{pmatrix}
0 & * & 0 \\
* & 0 & * \\
0 & * & *
\end{pmatrix}$
& 
$\begin{pmatrix}
0 & 0 & * \\
0 & * & * \\
* & * & 0
\end{pmatrix}$ \\ \hline
    \end{tabular}
    \caption{Yukawa matrices for $M=5$ with the flavor structure (i,i). The asterisk symbols "$*$" denote non-vanishing elements.}
    \label{tab:Y-M=5-1}
\end{table}

As another example, we replace the three generations of right-handed fermions with 
the three zero modes with the magnetic flux $M=5$, $\mathbf{Z}_2$ even, and SS phases $(1/2,0)$, while keeping the three generations of left-handed fermions as the previous one i.
That is, the three generations of right-handed fermions correspond to the following classes:
\begin{align}
\label{eq:flavor-assign-M=3-2}
    \text{ii} : ([g^{1/2}],[g^{3/2}],[g^{5/2}]),
\end{align}
while the three generations of left-handed fermions correspond to those given in Eq.~(\ref{eq:flavor-assign-M=3-1}).
In this case, the Higgs mode must correspond to one of $[g^{1/2}]$, $[g^{3/2}]$, and $[g^{5/2}]$ to obtain allowed Yukawa couplings.
Yukawa textures are shown in Table \ref{tab:Y-M=5-2}.
The texture in the last column is diagonal by permutation of rows and columns.
The other two textures correspond to the NNI pattern by permutation.

\begin{table}[H]
    \centering
    \begin{tabular}{|c|c|c|c|} \hline
      Flavor (left,right)  & Higgs $[g^{1/2}]$ & Higgs $[g^{3/2}]$ & Higgs $[g^{5/2}]$\\ \hline
\begin{tabular}{l}
(i,ii) 
\end{tabular} 
& 
$\begin{pmatrix}
* & 0 & 0 \\
* & * & 0 \\
0 & * & *
\end{pmatrix}$
& 
$\begin{pmatrix}
0 & * & 0 \\
* & 0 & * \\
* & * & 0
\end{pmatrix}$
& 
$\begin{pmatrix}
0 & 0 & * \\
0 & * & 0 \\
* & 0 & 0
\end{pmatrix}$ \\ \hline
    \end{tabular}
    \caption{Yukawa matrices for $M=5$ for patterns 3 in Table \ref{tab2}. The asterisk symbols "$*$" denote non-vanishing elements.}
    \label{tab:Y-M=5-2}
\end{table}

Similarly, we can analyze the case that both of left-handed and right-handed fermions correspond to the flavor structure ii in Eq.~(\ref{eq:flavor-assign-M=3-2}).
In this case, the Higgs mode must correspond to 
one of $[g^0]$, $[g^1]$, and $[g^2]$ to obtain allowed Yukawa couplings.
Yukawa textures are shown in Table \ref{tab:Y-M=5-3}.
The texture in the second column is diagonal.
The other two textures are the NNI pattern by permutation.

\begin{table}[H]
    \centering
    \begin{tabular}{|c|c|c|c|} \hline
      Flavor (left,right)  & Higgs $[g^{0}]$ & Higgs $[g^{1}]$ & Higgs $[g^{2}]$\\ \hline
\begin{tabular}{l}
(ii,ii)
\end{tabular}   
&  $\begin{pmatrix}
* & 0 & 0 \\
0 & * & 0 \\
0 & 0 & *
\end{pmatrix}$ & $\begin{pmatrix}
* & * & 0 \\
* & 0 & * \\
0 & * & 0
\end{pmatrix}$ 
& 
$\begin{pmatrix}
0 & * & * \\
* & * & 0 \\
* & 0 & 0
\end{pmatrix}$ \\ \hline
    \end{tabular}
    \caption{Yukawa matrices for $M=5$ for patterns 3 in Table \ref{tab2}. The asterisk symbols "$*$" denote non-vanishing elements.}
    \label{tab:Y-M=5-3}
\end{table}

Among 9 flavor models studied so far, three models lead to the diagonal texture, and the other 6 models lead to the NNI texture.

Similarly, we can analyze all the possible combinations of three generations of left-handed and right-handed fermions shown in Table \ref{tab:three_gen}  and the Higgs mode, which can couple.
Table \ref{tab:fluxes} shows the magnetic fluxes used to derive three generations of left-handed and right-handed fermions.
Totally, we have 167 combinations.
After identifying matrices related by row and column permutations, the resulting textures are classified into eight distinct types, which are shown in Table \ref{tab:texture_classification}.
In the fourth column, we denote the number of combinations by  $N_{\rm all}(=N_{\rm same}+N_{\rm diff})$, where $N_{\rm same}$ and $N_{\rm diff}$ denote the numbers of assignments in which the left- and right-handed quarks belong to the same and different classes, respectively. 
The ranks of Textures 5, 6, 8 are smaller than three.
In these textures, we can not realize all the masses of three generations.
However, their total rate is 9/167, which is rather small.
On the other hand, it is remarkable that  the NNI texture and the diagonal one are statistically favored.
The next one is Texture 1.
When we focus on the cases, where three generations of left-handed and right-handed fermions correspond to the same classes, the diagonal texture is the most favored.
On the other hand, when the flavor structure is asymmetric between left-handed and right-handed fermions, the NNI texture is the most favored.

\begin{table}[H]
    \centering
    \begin{tabular}{|c|c|} \hline
        SS phase & Magnetic fluxes $M$\\\hline
        Pattern 1 & 4,5 \\\hline
        Pattern 2 & 7,8\\\hline
        Pattern 3 & 5,6\\\hline
        Pattern 4 & 6,7 \\\hline
        Pattern 5 & 6,7\\\hline
        Pattern 6 & 5,6\\\hline
        Pattern 7 & 6,7\\\hline
        Pattern 8 & 6,7\\\hline
        Pattern 9 & 6,7\\\hline
        Pattern 10 & 5,6\\\hline
        Pattern 11 & 6,7\\\hline
        Pattern 12 & 5,6\\\hline
        Pattern 13 & 5,6\\\hline
        Pattern 14 & 5,6\\\hline
        Pattern 15 & 6,7\\\hline
    \end{tabular}
    \caption{Magnetic fluxes realizing three generations in each pattern.}
    \label{tab:fluxes}
\end{table}

In our multiplication rules, every class is self-conjugate.
That is their multiplications $[g^k][g^k]$ as well as $[g^{k+1/2}][g^{k+1/2}]$ always include $[g^0]$. 
If the left-handed and right-handed fermions correspond to the same classes and the Higgs corresponds to $[g^0]$, the diagonal entries in their Yukawa matrix are allowed.
That is the reason why the diagonal texture is most favorable when three generations of left-handed and right-handed fermions correspond to the same classes.

Textures 3, 5, 6, 7, and 8 can be derived by group-like symmetry if we choose  proper charges  for fermions and the Higgs modes.
The others can not be derived by group-like symmetry unless group-like symmetry is broken or multi-Higgs fields are introduced.
The ratio of the former textures to the latter is 78/89, which is roughly fifty-fifty.
Obviously, the NNI form is the most favored statistically when we focus on textures, which can not be derived by group-like symmetry.

\begin{table}[H]
\centering
\begin{tabular}{|c|c|c|c|c|}
\hline
Texture & Zero count & Pattern & Count & Percentage (\%) \\
\hline
1 & 3 &
$\begin{pmatrix}
* & * & 0 \\
* & 0 & * \\
0 & * & *
\end{pmatrix}$
& $20(=8+12)$ & $12.0~(10.5,~13.2)$ \\
\hline
2 (NNI) & 4 &
$\begin{pmatrix}
0 & * & 0 \\
* & 0 & * \\
0 & * & *
\end{pmatrix}$
& $66(=20+44)$ & $39.5~(26.3,~48.4)$ \\
\hline
3 & 4 &
$\begin{pmatrix}
* & * & 0 \\
* & * & 0 \\
0 & 0 & *
\end{pmatrix}$
& $8(=0+8)$ & $4.8~(0,~8.8)$ \\
\hline
4 & 5 &
$\begin{pmatrix}
* & 0 & 0 \\
0 & 0 & * \\
0 & * & *
\end{pmatrix}$
& $3(=2+1)$ & $1.8~(2.6,~1.1)$ \\
\hline
5 & 5 &
$\begin{pmatrix}
* & * & 0 \\
0 & 0 & * \\
0 & 0 & *
\end{pmatrix}$
& $6(=5+1)$ & $3.6~(6.6,~1.1)$ \\
\hline
6 & 5 &
$\begin{pmatrix}
0 & 0 & 0 \\
0 & * & * \\
0 & * & *
\end{pmatrix}$
& $1(=1+0)$ & $0.6~(1.3,~0)$ \\
\hline
7 & 6 &
$\begin{pmatrix}
* & 0 & 0 \\
0 & * & 0 \\
0 & 0 & *
\end{pmatrix}$
& $61(=38+23)$ & $36.5~(0.5,~25.3)$\\
\hline
8 & 7 &
$\begin{pmatrix}
0 & 0 & 0 \\
0 & 0 & * \\
0 & * & 0
\end{pmatrix}$
& $2(=2+0)$ &  $1.2~(2.6,~0)$\\
\hline
\end{tabular}
\caption{Classification of the Yukawa textures. 
In the fourth column, we denote the number of combinations by  $N_{\rm all}(=N_{\rm same}+N_{\rm diff})$, where $N_{\rm same}$ and $N_{\rm diff}$ denote the numbers of assignments in which the left- and right-handed quarks belong to the same and different classes, respectively. 
The percentages are given relatively by $N_{\rm all}/167~(N_{\rm same}/76,~N_{\rm diff}/91)$.}
\label{tab:texture_classification}
\end{table}

The above results show that the following Yukawa textures for the up-sector and down-sector of quarks:
\begin{align}
(a)\quad & Y_u=Y_d=
\begin{pmatrix}
    * & 0 & 0  \\
    0 & * & 0 \\
    0 & 0 & *
\end{pmatrix}, \\
(b)\quad & Y_u=
\begin{pmatrix}
    0& * & 0 \\
    * & 0 & * \\
    0 & * &*
\end{pmatrix}, \quad
Y_d=
\begin{pmatrix}
    * & 0 & 0  \\
    0 & * & 0 \\
    0 & 0 & *
\end{pmatrix}, \\
(c)\quad &
Y_u=
\begin{pmatrix}
    * & 0 & 0  \\
    0 & * & 0 \\
    0 & 0 & *
\end{pmatrix}, \quad 
 Y_d=
\begin{pmatrix}
    0& * & 0 \\
    * & 0 & * \\
    0 & * &*
\end{pmatrix}, \\
(d)\quad &
Y_u=Y_d=
\begin{pmatrix}
    0& * & 0 \\
    * & 0 & * \\
    0 & * &*
\end{pmatrix},
\end{align}
are favored statistically.

Since the texture $(a)$ can not realize flavor mixing, 
that is not realistic.
If the flavor assignments are the same between left-handed quarks and  right-handed quarks, as in an $SO(10)$-like flavor structure, the texture $(a)$ is statistically favored.
However, that is not realistic as mentioned above.
If only the down-sector right-handed quarks are different from the others as in an $SU(5)$-like flavor structure, the texture $(c)$ is statistically favored.
This can lead to realistic quark mass ratios and mixing angles.
This is an interesting result in this systematic analysis.

Texture 5, 6, and 8 themselves are not realistic, because their ranks are smaller than 3.
They are not statistically favored in the $T^2/\mathbf{Z}_2$ orbifold models.
These textures would be interesting when we consider multi-Higgs models, where one combines these textures with others such that they lead to three fermion masses and their mixing angles.
For example, the solution of the strong CP problem without axion was studied in Refs.~
\cite{Liang:2025dkm,Kobayashi:2025thd,Kobayashi:2025rpx}, where two or three Higgs fields are introduced.
One of the Higgs fields has  complex Yukawa coupling in a certain entry like Texture 9, such that it induces not the strong CP phase, but the weak CP phase.
Other Higgs fields have real Yukawa couplings, e.g., in the NNI form, in order to lead to three fermion masses and their mixing angles.

%%%%%%%%%%%%%%%%%%%%%%%%%%%%%%%%%%%%%%%%%%%%%%%%%%%%%%%%%%%%%%%%%

\section{Magnetized $T^2/\mathbf{Z}_4$ orbifold models}
\label{sec:Z4-orbifold}

Similarly, we can study Yukawa textures derived from other magnetized orbifold models.
The $\mathbf{Z}_M$ symmetry constrains strongly allowed couplings in magnetized $T^2$ models, where GCD of fluxes for left-handed and right-handed fermions and the Higgs mode is equal to $M$.
In magnetized $T^2/\mathbf{Z}_2$ orbifold models with vanishing SS phases, the coupling selection rules becomes weaker and are controlled by $\mathbf{Z}_2$ gauging of $\mathbf{Z}_M$ symmetry. 
That corresponds to multiplication rules of $\mathbf{Z}_2$-invariant conjugacy classes of $D_M$.
We find similar selection rules in 
magnetized $T^2/\mathbf{Z}_2$ orbifold models with non-vanishing SS phases.
On the other hand, zero-mode wave functions in other $T^2/\mathbf{Z}_N$ orbifolds correspond to linear combinations of wave functions with various $\mathbf{Z}_M$ charges.
For such linear combinations on other orbifolds, selection rules would be trivial and all of entries in the Yukawa matrix would be allowed as 
\begin{align}
\label{eq:trivial-textures}
    Y=
    \begin{pmatrix}
        * & * & * \\
        * & * & * \\
        * & * & * \\
    \end{pmatrix}.
\end{align}

Here, we fucus on  magnetized $T^2/\mathbf{Z}_4$ orbifold models.\footnote{Similarly we can study other orbifold models.}
The selection rules in magnetized $T^2/\mathbf{Z}_2$ orbifold models such as $\mathbf{Z}_2$ gauging of $\mathbf{Z}_M$ symmetries would be further violated by the $\mathbf{Z}_4$ twist of the $T^2/\mathbf{Z}_4$ orbifold.
Any selection rules originating from the $\mathbf{Z}_M$ symmetries may not remain in magnetized $T^2/\mathbf{Z}_4$ orbifold models like Eq.~(\ref{eq:trivial-textures}).
Here, we examine whether any selection rules originating from the $\mathbf{Z}_M$ symmetries remain or not.
We study the possible textures in magnetized $T^2/\mathbf{Z}_4$ orbifold models.

\subsection{Zero-mode wave functions}

Here, we review the $T^2/\mathbf{Z}_4$ orbifold models \cite{Abe:2013bca,Abe:2014noa,Kobayashi:2017dyu}. 
As reviewed in section \ref{sec:orbifold},
zero-mode wavefunctions on $T^2$ with the magnetic flux $M$ and vanishing SS phases can be written by the following basis of functions:
\begin{equation}
\label{eq:basis1}
\psi^{j,M}(z,\tau) = \mathcal{N} \cdot e^{i \pi M z  \mathrm{Im}{z}/ \mathrm{Im}\tau } \cdot \vartheta \left[
\begin{array}{c}
\frac{j}{M} \\
0
\end{array}
\right] ( M z, M \tau) .
\end{equation}
The same zero-mode wave functions can be expanded by using another basis of wave functions as follows \cite{Cremades:2004wa},
\begin{equation}
\label{eq:basis2}
\chi^{j,M}(\tau,z) = \frac{\mathcal{N}}{\sqrt{M}} \cdot e^{i \pi M z  \mathrm{Im}{z}/ \mathrm{Im}{\tau} } \cdot \vartheta \left[
\begin{array}{c}
0 \\
\frac{j}{M}
\end{array}
\right] ( z, \tau/M ).
\end{equation}
These are related by $S$ transformation: $(z,\tau)\xrightarrow{}(z/\tau,-1/\tau)$,
\begin{align}
    S\cdot\chi^{j,M}(z,\tau)=\chi^{j,M}\left(\dfrac{z}{\tau},-\dfrac{1}{\tau}\right)=\psi^{j,M}(z,\tau).
\end{align}
For the zero-mode wavefunctions to be well-defined on $T^2/\mathbf{Z}_4$, the $\mathbf{Z}_4$ twist must be represented linearly by wave functions. 
The $\mathbf{Z}_4$ twist can be understood by $S^{-1}$ transformation with $\tau=i$: $(z,\tau=i)\xrightarrow{}(-z/i,-1/i)=(iz,i)$. Thus, under such $\mathbf{Z}_4$ twist, wavefunctions transform
\begin{eqnarray}
\psi^{j,M}(z,\tau=i) \rightarrow & & \psi^{j,M}(iz,-1/\tau = i )   \nonumber \\
                             & &= \chi^{j,M}(z,\tau = i) \\ 
& &= C^j_{k,M}\psi^{k,M}(z,\tau=i),    \nonumber 
\end{eqnarray}
 where \begin{equation}
C^{j}_{k,M} = \frac{1}{\sqrt{M}} e^{2\pi i \frac{jk}{M}}  .
\end{equation}

When the $\mathbf{Z}_4$ twist is applied twice, the matrix $C^{j}_{k,M}$ satisfies the following equation,
\begin{equation}
\sum_k C^{j}_{k,M}C^{k}_{l,M} = \frac{1}{M} \sum_k e^{2 \pi i (j+l)k/M} = \delta_{(j+l),~nM}.
\end{equation}
Hence, we find 
\begin{align}
    \sum_{l}(C^2)^{j}_{l,M}\psi^{l,M}=\psi^{nM-j,M}\equiv\psi^{M-j,M}\quad (\mathrm{mod}\ M).
\end{align}
Then, the zero-mode wavefunctions on $T^2/\mathbf{Z}_4$ can be expressed as the following linear combinations:
\begin{equation}
\frac12 \left(   \psi^{j,M}(z,i) +\gamma^{-1} \sum_k C^j_{k,M}\psi^{k,M}(z,i) + \gamma^{-2} \psi^{M-j,M}(z,i)  + \gamma^{-3}  \sum_k C^{M-j}_{k,M}\psi^{k,M}(z,i)  \right),
\end{equation}
whose coefficients are the $\mathbf{Z}_4$ eigenvalue $\gamma=\pm1,\pm i$.

We now express the zero-mode wavefunctions on $T^2/\mathbf{Z}_4$ 
as linear combinations of the wavefunctions on $T^2$, 
\begin{align}
    \Psi_{T^2/\mathbf{Z}_4}(z)=\sum_{j=0}^{M-1}a_j\psi^{j,M}(z),
\end{align}
where $a_j\ (j=0,\cdots, M-1)$ denote the expansion coefficients.
When $\Psi_{T^2/\mathbf{Z}_4}(z)$ is a $\mathbf{Z}_4$ eigenstate with eigenvalue $\gamma$, we can derive the following equation:
\begin{align}
\sum_{j=0}^{M-1} a_j
\sum_{k=0}^{M-1} C^{j}_{k,M}\psi^{k,M}(z)
=
\sum_{k=0}^{M-1}
\left(
\sum_{j=0}^{M-1} C^{j}_{k,M}a_j
\right)
\psi^{k,M}(z)=\sum_{k=0}^{M-1}
\gamma a_k\psi^{k,M}(z).
\end{align}
Since the wavefunctions $\psi^{k,M}(z)$ are linearly independent, this yields the following eigenvalue equation for $\gamma$:
\begin{equation}
\sum_{j=0}^{M-1}C^{j}_{k,M}a_j=\gamma a_k,\qquad
(k=0,\ldots,M-1).
\label{eq:eigen}
\end{equation}
Multiplying both sides of Eq.\eqref{eq:eigen} by $C^{k}_{\ell,M}$ and summing over $k$ yields
\begin{equation}
\sum_{j=0}^{M-1}
\left(C^2\right)^{j}_{l,M}a_j=\gamma^2 a_{\ell}.
\end{equation}
Using Eq.\eqref{eq:eigen}, it becomes  
\begin{align}
\sum_{j=0}^{M-1}
\delta_{j+l,\,0}
a_j=\gamma^2a_l\quad (\mathrm{mod}\ M), 
\end{align}
and equivalently, 

\begin{align}
    a_{M-k}=\gamma^2 a_k\quad (\mathrm{mod}\ M).
\end{align}
Therefore, for $M=2m$, this condition requires
$\bm{a}=(a_0,a_1,\cdots)$ to have the following form:
\begin{align}
    &\gamma=\pm 1: (a_0,a_1,\cdots, a_{m-1},a_m,a_{m-1},\cdots, a_1),\label{eq:g1}\\
    &\gamma=\pm i: (0, a_1, \cdots, a_{m-1},0,-a_{m-1},\cdots, -a_1).
\end{align}
Similarly, for $M=2m+1$, the eigenvector $\bm{a}$ must have the following form:
\begin{align}
    &\gamma=\pm 1: (a_0,a_1,\cdots, a_m,a_m,\cdots, a_1),\\
    &\gamma=\pm i: (0, a_1, \cdots, a_m,-a_m,\cdots, -a_1)\label{eq:gi}.
\end{align}
Substituting Eqs.\eqref{eq:g1}$-$\eqref{eq:gi} into Eq.\eqref{eq:eigen} yields the following system of simultaneous equations. For $M=2m$, we obtain
\begin{align}
    &\gamma=\pm 1: a_0+(-1)^ka_m+2\sum_{j=1}^{m-1}a_j\cos\left(\pi\frac{ jk}{m}\right)=\sqrt {2m}\,\gamma a_k\qquad (k=0,\cdots, m),\label{eq:simul}\\
    &\gamma=\pm i:2i\sum_{j=1}^{m-1}
a_j\sin\left(\pi\frac{ jk}{m}\right)=\sqrt {2m}\,\gamma a_k
\qquad (k=1,\cdots, m-1).
\end{align}
For $M=2m+1$,  we obtain 
\begin{align}
    &\gamma=\pm 1:a_0+2\sum_{j=1}^{m}a_j\cos\left(\frac{2\pi jk}{2m+1}\right)=\sqrt {2m+1}\,\gamma a_k,\qquad (k=0,\ldots,m),\\
    &\gamma=\pm i:2i\sum_{j=1}^{m}a_j\sin\left(\frac{2\pi jk}{2m+1}\right)=\sqrt {2m+1}\,\gamma a_k,\qquad (k=1,\ldots,m).\label{eq:simul2}
\end{align}

For the specific case with magnetic flux $M=8$, let us determine the eigenvector $\bm{a}$ satisfying Eq.\eqref{eq:eigen}. In this case, by substituting $m=4$ into Eq.\eqref{eq:simul}$-$\eqref{eq:simul2} and solving the resulting system of simultaneous equations, the eigenvector $\bm{a}$ for each eigenvalue is obtained as follows:
\begin{align}
\bm{a}_{\gamma=+1}=\Bigg(&\frac{a_1}{\sqrt2}+(1+\sqrt2)a_2
+\frac{a_3}{\sqrt2},
\,a_1,a_2,a_3,-\frac{a_1}{\sqrt2}+(1+\sqrt2)a_2-\frac{a_3}{\sqrt2},\,a_3,a_2,a_1\Bigg),\\
\bm{a}_{\gamma=-1}=\Bigl(&-\sqrt{2}\,a_1+(1-\sqrt{2})a_2,\,
a_1,\,a_2,\,a_1,\sqrt{2}\,a_1+(1-\sqrt{2})a_2,\,a_1,\,a_2,\,a_1\Bigr),\\
\bm{a}_{\gamma=+i}=\Bigl(&0,\,a_1,\,a_2,\,a_1-\sqrt{2}\,a_2,\,0,-(a_1-\sqrt{2}\,a_2),\,-a_2,\,-a_1\Bigr),\\
\bm{a}_{\gamma=-i}=\Bigl(&0,\,a_1,\,-\sqrt{2}\,a_1,\,-a_1,\,
0,\,a_1,\,\sqrt{2}\,a_1,\,-a_1\Bigr).
\end{align}

For example, for the eigenvalue $\gamma=+1$, the corresponding eigenspace is three-dimensional, and 
we choose the basis of wave functions as follows,

\begin{align}
\Psi^{0,8}_{{T^2/\mathbf Z_4},1}(z)={}&\frac{1}{2}\Bigl(
 \psi^{1,8}_{T^2}
-\psi^{3,8}_{T^2}
-\psi^{5,8}_{T^2}
+\psi^{7,8}_{T^2}
\Bigr),\\[1ex]
\Psi^{1,8}_{{T^2/\mathbf Z_4},1}(z)
={}&\frac{1}{2\sqrt2}\Bigl(
 \sqrt2\,\psi^{0,8}_{T^2}
+\psi^{1,8}_{T^2}
+\psi^{3,8}_{T^2}
-\sqrt2\,\psi^{4,8}_{T^2}
+\psi^{5,8}_{T^2}
+\psi^{7,8}_{T^2}
\Bigr),\\[1ex]
\Psi^{2,8}_{{T^2/\mathbf Z_4},1}(z)
={}&\frac{1}{2\sqrt{2+\sqrt2}}\Bigl(
 (1+\sqrt2)\psi^{0,8}_{T^2}
+\psi^{2,8}_{T^2}
 +(1+\sqrt2)\psi^{4,8}_{T^2}
+\psi^{6,8}_{T^2}
\Bigr).
\end{align}
Note that $\Psi^{0,8}_{{T^2/\mathbf Z_4},1}(z)$ includes $\psi^{j,8}_{T^2}$ only with  $j=$ odd, while $\Psi^{2,8}_{{T^2/\mathbf Z_4},1}(z)$ includes $\psi^{j,8}_{T^2}$ only with  $j=$ even.
On the other hand, $\Psi^{1,8}_{{T^2/\mathbf Z_4},1}(z)$ includes $\psi^{j,8}_{T^2}$ with $j=0, 1, 3, 4, 5, 7$.
Then, we can expect that some selection rules, which originate from the $\mathbf{Z}_8$ symmetry, may remain.
Indeed, in the next subsection we will show that we can realize nontrivial textures.
Similarly, we choose the basis of zero mode wave functions such that we can derive more zero textures.
The numbers of zero modes for each magnetic flux $|M|$ and each $\mathbf{Z}_4$ eigenvalue for vanishing SS phase are summarized in Table \ref{tab:zero-modes}\ \cite{Abe:2013bca,Abe:2014noa}.

\begin{table}[H]
\centering
\begin{tabular}{|cc|ccccccccccccccccc|}
\hline
\multicolumn{2}{|c|}{$|M|$} & 1 & 2 & 3 & 4 & 5 & 6 & 7 & 8 & 9 & 10 & 11 & 12 & 13 & 14 & 15 & 16 & 17 \\ \hline
\multicolumn{1}{|c|}{\multirow{4}{*}{$\gamma$}} 
& $1$ & 1 & 1 & 1 & 2 & 2 & 2 & 2 & 3 & 3 & 3 & 3 & 4 & 4 & 4 & 4 & 5 & 5 \\
\multicolumn{1}{|c|}{} & $i$ & 0 & 0 & 1 & 1 & 1 & 1 & 2 & 2 & 2 & 2 & 3 & 3 & 3 & 3 & 4 & 4 & 4 \\
\multicolumn{1}{|c|}{} & $-1$ & 0 & 1 & 1 & 1 & 1 & 2 & 2 & 2 & 2 & 3 & 3 & 3 & 3 & 4 & 4 & 4 & 4 \\
\multicolumn{1}{|c|}{} & $-i$ & 0 & 0 & 0 & 0 & 1 & 1 & 1 & 1 & 2 & 2 & 2 & 2 & 3 & 3 & 3 & 3 & 4 \\ \hline
\end{tabular}
\caption{The number of linearly independent zero-mode eigenstates $\Psi_{T^2/\mathbf{Z}_4,\gamma}(z)$ for $(\alpha_1,\alpha_\tau)=(0,0)$ on $T^2/\mathbf{Z}_4$.}
\label{tab:zero-modes}
\end{table}

Similarly, there are zero-mode wavefunctions with the SS phase $(1/2,1/2)$ on $T^2/\mathbf{Z}_4$. 
Note that only the possible SS phase is $(1/2,1/2)$ in magnetized $T^2/\mathbf{Z}_4$ orbifold models \cite{Abe:2013bca}.
Then, the degeneracies of the zero-mode wavefunctions are summarized in Table \ref{tab:withSSphase} \cite{Abe:2013bca,Abe:2014noa}.

\begin{table}[H]
  \centering
  \setlength{\tabcolsep}{4.5pt}
  \renewcommand{\arraystretch}{1.15}
  \begin{tabular}{|c|c|*{16}{c}|}
    \hline
    \multicolumn{2}{|c|}{$|M|$}
      & 1 & 2 & 3 & 4 & 5 & 6 & 7 & 8
      & 9 & 10 & 11 & 12 & 13 & 14 & 15 & 16 \\
    \hline
    \multirow{4}{*}{$\gamma$}
      & $1$ & 0 & 1 & 1 & 1 & 1 & 2 & 2 & 2
             & 2 & 3 & 3 & 3 & 3 & 4 & 4 & 4 \\
      & $i$ & 1 & 1 & 1 & 1 & 2 & 2 & 2 & 2
             & 3 & 3 & 3 & 3 & 4 & 4 & 4 & 4 \\
      & $-1$ & 0 & 0 & 0 & 1 & 1 & 1 & 1 & 2
             & 2 & 2 & 2 & 3 & 3 & 3 & 3 & 4 \\
      & $-i$ & 0 & 0 & 1 & 1 & 1 & 1 & 2 & 2
             & 2 & 2 & 3 & 3 & 3 & 3 & 4 & 4 \\
    \hline
  \end{tabular}
  \caption{The number of linearly independent zero-mode eigenstates $\Psi_{T^2/\mathbf{Z}_4,\gamma}(z)$ for $(\alpha_1,\alpha_\tau)=(1/2,1/2)$ on $T^2/\mathbf{Z}_4$.}
  \label{tab:withSSphase}
\end{table}

\subsection{Yukawa textures on $T^2/\mathbf{Z}_4$ from selection rule}
Now, we study Yukawa textures that give rise to a three-generation fermion structure.

\subsubsection{Vanishing SS phases}

First, we consider the cases with vanishing SS phases.
Since the degeneracy of zero modes corresponds to the number of generations in the four-dimensional effective theory, the combinations of magnetic fluxes and $\mathbf{Z}_4$ eigenvalues that yield three generations are summarized in Table \ref{tab:combinations_M}. Note that, in this analysis, we focus on the case in which the three magnetic fluxes take the form $(M,M,2M)$.
Other combinations of three magnetic fluxes satisfy the gauge invariance condition $M_1+M_2=M_3$.
However, if their greatest common division is equal to 1, the selection rules become trivial, and all entries of the Yukawa texture are allowed as Eq.~(\ref{eq:trivial-textures}).

\begin{table}[H]
    \centering
    \begin{tabular}{|c|c|}\hline
       $(M_1,M_2,M_3)$   & $(\gamma_1,\gamma_2,\gamma_3^{*})$\\ \hline
        $(8,8,16)$ & $(1,1,1)$\\\hline
        $(9,9,18)$ & $(1,1,1)$\\\hline
        $(10,10,20)$ & $(1,1,1),(-1,-1,1)$,$(1,-1,-1)$\\\hline
        $(11,11,22)$ & $(1,1,1),(-1,-1,1)$,$(1,-1,-1)$,$(1,i,-i)$,$(i,i,-1)$,$(i,-1,i)$\\\hline
        $(12,12,24)$ & $(-1,-1,1)$,$(i,i,-1)$,$(i,-1,i)$\\\hline
        $(13,13,26)$ & $(i,-i,1)$,$(-1,-1,1)$,$(i,i,-1)$,$(i,-1,i)$,$(-1,-i,-i)$,$(-i,-i,-1)$\\\hline
        $(14,14,28)$ & $(i,i,-1)$,$(i,-i,1)$,$(-i,-i,-1)$\\\hline
        $(15,15,30)$ & $(-i,-i,-1)$\\\hline
        $(16,16,32)$ &$(-i,-i,-1)$\\ \hline
    \end{tabular}
    \caption{Combinations of magnetic fluxes $(M,M,2M)$ and $\mathbf{Z}_4$ eigenvalues $\gamma$ that yield three generations. Here, we consider vanishing SS phase.}
    \label{tab:combinations_M}
\end{table}

In each combination of $(M_1,M_2,M_3)$, there are various candidates for the Higgs wave functions, and derived textures depend on it.
Table \ref{tab:texture00} shows the texture derived from each of the Higgs wave functions for $(M_1,M_2,M_3)$. 
In the cases other than shown in Table \ref{tab:texture00}, we obtain the trivial result, where all entries in Yukawa matrix are allowed as Eq.~(\ref{eq:trivial-textures}).
Table \ref{tab:texture00} includes non-trivial textures, which cannot be realized by group-theoretical symmetries.

\begin{table}[H]
    \centering
    \begin{tabular}{|c|c|c|c|c|c|c|c|}
        \hline
         $(M_1,M_2,M_3)$ & $(\gamma_1,\gamma_2,\gamma_3^{*})$ & \multicolumn{6}{c|}{Yukawa textures} \\
        \hline
         (8,8,16) & (1,1,1) & 
            $\begin{pmatrix}
                0 & * & * \\
                * & * & * \\
                * & * & 0
            \end{pmatrix}$& 
            $\begin{pmatrix}
                * & * & * \\
                * & * & * \\
                * & * & *
            \end{pmatrix}$& 
            $\begin{pmatrix}
                * & * & * \\
                * & * & * \\
                * & * & *
            \end{pmatrix}$& 
            $\begin{pmatrix}
                * & * & 0 \\
                * & * & * \\
                0 & * & *
            \end{pmatrix}$&
            $\begin{pmatrix}
                * & * & 0 \\
                * & * & * \\
                0 & * & *
            \end{pmatrix}$ & \\
        \hline
         (9,9,18) & (1,1,1) &
            $\begin{pmatrix}
                * & * & * \\
                * & 0 & * \\
                * & * & *
            \end{pmatrix}$& 
            $\begin{pmatrix}
                * & * & * \\
                * & 0 & * \\
                * & * & *
            \end{pmatrix}$& 
            $\begin{pmatrix}
                * & 0 & * \\
                0 & * & * \\
                * & * & *
            \end{pmatrix}$& 
            $\begin{pmatrix}
                * & * & * \\
                * & * & * \\
                * & * & *
            \end{pmatrix}$&
            $\begin{pmatrix}
                * & * & * \\
                * & * & * \\
                * & * & *
            \end{pmatrix}$ & \\
        \hline
         (12,12,24) &$(i,i,-1)$&
            $\begin{pmatrix}
                * & * & * \\
                * & * & * \\
                * & * & *
            \end{pmatrix}$& 
            $\begin{pmatrix}
                * & * & * \\
                * & * & * \\
                * & * & *
            \end{pmatrix}$& 
            $\begin{pmatrix}
                * & * & * \\
                * & * & * \\
                * & * & *
            \end{pmatrix}$& 
            $\begin{pmatrix}
                * & * & * \\
                * & * & * \\
                * & * & *
            \end{pmatrix}$&
            $\begin{pmatrix}
                * & 0 & * \\
                0 & * & * \\
                * & * & *
            \end{pmatrix}$&
            $\begin{pmatrix}
                * & 0 & * \\
                0 & * & * \\
                * & * & *
            \end{pmatrix}$
            \\
        \hline
    \end{tabular}
    \caption{Three-generation Yukawa textures with vanishing entries when the SS phases are $(\alpha_1,\alpha_\tau)=(0,0)$ in all three sectors.
}
    \label{tab:texture00}
\end{table}

\subsubsection{Non-vanishing SS phases in left-handed and right-handed fermions}

Similarly, we can analyze the cases, where 
both the left-handed and right-handed fermions have non-vanishing SS phases,   $(1/2,1/2)$. 
In this case, the Higgs mode must have  vanishing SS phases $(0,0)$ because of gauge invariance.
In this case, from Table \ref{tab:withSSphase}, the combinations of magnetic fluxes and $\mathbf{Z}_4$ eigenvalues that yield three generations are summarized in Table \ref{tab:three-generations-with-ss-phase}.

\begin{table}[H]
    \centering
    \begin{tabular}{|c|c|}\hline
    $(M_1,M_2,M_3)$& $(\gamma_1,\gamma_2,\gamma_3^
    {*})$ \\\hline

 $(9,9,18)$
& $(i,i,-1)$ \\
\hline

$(10,10,20)$
& $(1,1,1)$,
  $(1,i,-i)$,
  $(i,i,-1)$ \\
\hline

\multirow{2}{*}{$(11,11,22)$}
& $(1,1,1)$,
  $(1,i,-i)$,
  $(1,-i,i)$, \\
& $(i,i,-1)$,
  $(i,-i,1)$,
  $(-i,-i,-1)$ \\
\hline

\multirow{3}{*}{$(12,12,24)$}
& $(1,1,1)$,
  $(1,i,-i)$,
  $(1,-1,-1)$, \\
& $(1,-i,i)$,
  $(i,i,-1)$,
  $(i,-1,i)$, \\
& $(i,-i,1)$,
  $(-1,-1,1)$,
  $(-1,-i,-i)$,
  $(-i,-i,-1)$ \\
\hline

\multirow{2}{*}{$(13,13,26)$}
& $(1,1,1)$,
  $(1,-1,-1)$,
  $(1,-i,i)$,\\
& $(-1,-1,1)$,
  $(-1,-i,-i)$,
  $(-i,-i,-1)$ \\
\hline

$(14,14,28)$
& $(-1,-1,1)$,
  $(-1,-i,-i)$,
  $(-i,-i,-1)$ \\
\hline

$(15,15,30)$
& $(-1,-1,1)$ \\
\hline
\end{tabular}

  \caption{
    Combinations of magnetic fluxes $(M,M,2M)$ and $\mathbf{Z}_4$
    eigenvalues $\gamma$ yielding three generations for the SS phases
    $(\alpha_1^{(1)},\alpha_{\tau}^{(1)})
    =
    (\alpha_1^{(2)},\alpha_{\tau}^{(2)})
    =
    (1/2,1/2)$.
    Consequently, the SS phase in the third sector is
    $(\alpha_1^{(3)},\alpha_{\tau}^{(3)})
    =(0,0)\pmod{1}$.
  }
  \label{tab:three-generations-with-ss-phase}
\end{table}

Based on Table \ref{tab:three-generations-with-ss-phase}, the resulting Yukawa textures containing zero entries are summarized in Table\ref{tab:texture1212} depending on the Higgs wave function. 
All Yukawa textures obtained from combinations of magnetic fluxes and eigenvalues other than those listed in Table \ref{tab:texture1212} have all their entries allowed as Eq.~(\ref{eq:trivial-textures}).

\begin{table}[H]
    \centering
    \begin{tabular}{|c|c|c|c|c|c|c|}
        \hline
         $(M_1,M_2,M_3)$ & $(\gamma_1,\gamma_2,\gamma_3^{*})$ & \multicolumn{5}{c|}{Yukawa textures} \\
        \hline         
         (9,9,18) & $(i,i,-1)$ &
            $\begin{pmatrix}
                0 & * & * \\
                * & * & * \\
                * & * & *
            \end{pmatrix}$& 
            $\begin{pmatrix}
                0 & * & * \\
                * & * & * \\
                * & * & *
            \end{pmatrix}$& 
            $\begin{pmatrix}
                * & 0 & * \\
                0 & * & * \\
                * & * & *
            \end{pmatrix}$& 
            $\begin{pmatrix}
                * & * & * \\
                * & * & * \\
                * & * & *
            \end{pmatrix}$&
            $\begin{pmatrix}
                * & * & * \\
                * & * & * \\
                * & * & *
            \end{pmatrix}$\\\hline
    \end{tabular}
    \caption{Three-generation Yukawa textures with vanishing entries when both the first and second sectors have the SS phases $(\alpha_1,\alpha_\tau)=(1/2,1/2)$.}
    \label{tab:texture1212}
\end{table}

\subsubsection{Non-vanishing SS phases for left-handed fermions and the Higgs mode}

Finally, we consider the remaining combination of SS phases, namely, the case, in which the left-handed fermions, right-handed fermions, and the Higgs mode have the SS phases $(1/2,1/2)$, $(0,0)$, and $(1/2,1/2)$, respectively. In this case, from Tables  \ref{tab:zero-modes} and \ref{tab:withSSphase}, the combinations of magnetic fluxes and $\mathbf{Z}_4$ eigenvalues that yield three generations are summarized in Table \ref{tab:three-generations-half-zero-half}.
It is obvious that the case, in which the left-handed fermions, right-handed fermions, and the Higgs mode have the SS phases $(0,0)$, $(1/2,1/2)$,  and $(1/2,1/2)$, respectively, is obtained by replacing the rows and columns.

\begin{table}[H]
    \centering
    \begin{tabular}{|c|c|}\hline
    $(M_1,M_2,M_3)$& $(\gamma_1,\gamma_2,\gamma_3^
    {*})$ \\\hline

$(9,9,18)$
& $(i,1,-i)$ \\
\hline

$(10,10,20)$
& $(1,1,1)$,
  $(1,-1,-1)$,  $(i,1,-i)$,
  $(i,-1,i)$ \\
\hline

\multirow{2}{*}{$(11,11,22)$}
& $(1,1,1)$,
  $(1,i,-i)$,
  $(1,-1,-1)$, 
$(i,1,-i)$, 
  $(i,i,-1)$, \\
&  $(i,-1,i)$,
 $(-i,1,i)$,
  $(-i,i,1)$,
  $(-i,-1,-i)$ \\
\hline

\multirow{2}{*}{$(12,12,24)$}
& $(1,i,-i)$,
  $(1,-1,-1)$,
  $(i,i,-1)$, 
 $(i,-1,i)$,\\
 & $(-1,i,i)$,
  $(-1,-1,1)$, 
 $(-i,i,1)$,
  $(-i,-1,-i)$ \\
\hline

\multirow{2}{*}{$(13,13,26)$}
& $(1,i,-i)$,
  $(1,-1,-1)$,
  $(1,-i,i)$, 
 $(-1,i,i)$,
  $(-1,-1,1)$,\\
 & $(-1,-i,-i)$, 
 $(-i,i,1)$,
  $(-i,-1,-i)$,
  $(-i,-i,-1)$ \\
\hline

$(14,14,28)$
& $(-1,i,i)$,
  $(-1,-i,-i)$, 
 $(-i,i,1)$,
  $(-i,-i,-1)$ \\
\hline

$(15,15,30)$
& $(-1,-i,-i)$ \\
\hline
\end{tabular}

  \caption{
    Combinations of magnetic fluxes $M$ and $\mathbf{Z}_4$
    eigenvalues $\gamma$ yielding three generations for the SS phases
    $(\alpha_1^{(1)},\alpha_{\tau}^{(1)})
    =(1/2,1/2)$ and
    $(\alpha_1^{(2)},\alpha_{\tau}^{(2)})
    =(0,0)$.
    Consequently, the SS phase in the third sector is
    $(\alpha_1^{(3)},\alpha_{\tau}^{(3)})
    =(1/2,1/2)\pmod{1}$.
  }
  \label{tab:three-generations-half-zero-half}
\end{table}

Based on Table \ref{tab:three-generations-half-zero-half}, the resulting Yukawa textures containing zero entries are summarized in Table\ref{tab:texture1200}, depending on the Higgs wave function. 
All Yukawa textures obtained from combinations of magnetic fluxes and eigenvalues other than those listed in Table \ref{tab:texture1200} have all their entries allowed as Eq.~(\ref{eq:trivial-textures}).

\begin{table}[H]
    \centering
    \begin{tabular}{|c|c|c|c|c|c|c|}
        \hline
         $(M_1,M_2,M_3)$ & $(\gamma_1,\gamma_2,\gamma_3^{*})$ & \multicolumn{5}{c|}{Yukawa textures} \\
        \hline         
         (9,9,18) & $(i,i,-1)$ &
            $\begin{pmatrix}
                0 & * & * \\
                * & 0 & * \\
                * & * & *
            \end{pmatrix}$& 
            $\begin{pmatrix}
                * & 0 & * \\
                * & * & * \\
                * & * & *
            \end{pmatrix}$& 
            $\begin{pmatrix}
                * & * & * \\
                * & * & * \\
                * & * & *
            \end{pmatrix}$& 
            $\begin{pmatrix}
                * & 0 & * \\
                * & * & * \\
                * & * & *
            \end{pmatrix}$&
            $\begin{pmatrix}
                * & * & * \\
                * & * & * \\
                * & * & *
            \end{pmatrix}$\\\hline
    \end{tabular}
    \caption{Three-generation Yukawa textures with vanishing entries when both the first and third sectors have the SS phases $(\alpha_1,\alpha_\tau)=(1/2,1/2)$.}
    \label{tab:texture1200}
\end{table}

To summarize, we can derive the following three textures:
\begin{align}
    \begin{pmatrix}
        * & * & * \\
        * & * & * \\
        * & * & * \\
    \end{pmatrix}, \qquad 
     \begin{pmatrix}
        0 & * & * \\
        * & * & * \\
        * & * & * \\
    \end{pmatrix}, \qquad 
     \begin{pmatrix}
        0 & * & * \\
        * & 0 & * \\
        * & * & * \\
    \end{pmatrix}, 
\end{align}
in magnetized $T^2/\mathbf{Z}_4$ orbifold models with and without SS phases.
The last two textures are nontrivial and cannot be derived by group-theoretical symmetries, 
although many cases lead to the first trivial texture.

\section{Additional cancellations by explicit calculations of Yukawa couplings}

So far, we have studied the coupling selection rules originated from the $\mathbf{Z}_M$ symmetry.
Magnetized $T^2$ models have the exact $\mathbf{Z}_M$ symmetry, which control the Yukawa textures as well as other couplings.
On $T^2/\mathbf{Z}_N$ orbifolds, the exact $\mathbf{Z}_M$ symmetry is violated, but some rules remain.
The coupling selection rules in magnetized $T^2/\mathbf{Z}_2$ orbifolds without SS phases are controlled by $\mathbf{Z}_2$ gauging of $\mathbf{Z}_M$ symmetries as shown in Refs.~\cite{Kobayashi:2024yqq,Kobayashi:2024cvp}.
Magnetized $T^2/\mathbf{Z}_2$ orbifold models with SS phases have similar selection rules, and they lead to nontrivial textures.
In magnetized $T^2/\mathbf{Z}_4$ orbifold models with fluxes $(M,M,2M)$, some selection rules, which originate from $\mathbf{Z}_M$ symmetries, remain, and they lead to nontrivial textures, although we obtain trivial textures like Eq.~(\ref{eq:trivial-textures}) in most of magnetized $T^2/\mathbf{Z}_4$ orbifold models. 
In this section, we show that further zero textures appear in the Yukawa textures for the flux configuration $(M,M,2M)$ by explicit evaluation of Yukawa couplings.

We consider the Yukawa coupling on $T^2$ with vanishing SS phases again. From Eq.\eqref{eq:yukawa-coupling}, Yukawa coupling without the normalization factor is given by

\begin{align}
\sum_{m=0}^{2M-1}
\vartheta
\begin{bmatrix}
\dfrac{i-j+Mm}{2M^2}\\0
\end{bmatrix}
\left(0,
2M^3\tau\right)\cdot\delta_{i+j-a,\,2M\ell -Mm}.
\label{eq:reduced}
\end{align}
The condition due to the Kronecker delta is given by
\begin{align}
    i+j-a+Mm=0\quad (\mathrm{mod}\ 2M).
\end{align}
We decompose $m=0,1,\cdots,2M-1$ as $m=2r+\epsilon$, where $r=0,1,\ldots,M-1$, $\epsilon=0,1$. Substituting this decomposition into the condition imposed by the Kronecker delta, we obtain
\begin{align}
i+j-a+Mm
&=i+j-a+M(2r+\epsilon)\nonumber\\
&=i+j-a+2Mr+M\epsilon\nonumber\\
&= i+j-a+M\epsilon\pmod{2M}.
\end{align}
Hence, when $i+j-a=0$, we have $\epsilon=0$, so that only even values of $m$ survive. By contrast, when $i+j-a=M$, we have $\epsilon=1$, and only odd values of $m$ survive. We then define $b=i-j+M\epsilon\quad(\mathrm{mod}\ 2M).$
Then, provided that the coupling in Eq.\eqref{eq:reduced} satisfies the selection rule, it can be written as follows,
\begin{align}
  \Theta_b(\tau)
  & =
  \sum_{r=0}^{M-1}
  \vartheta
  \begin{bmatrix}
      \dfrac{b+2Mr}{2M^2} \\
      0
  \end{bmatrix}
  \left(0,2M^3\tau\right).
  \label{eq:Theta-definition}
\end{align}
It can be further rewritten in the following form:
\begin{align}
  \Theta_b(\tau)
  ={}&
  \sum_{r=0}^{M-1}\sum_{n\in\mathbf{Z}}
  \exp\!\left[
    \pi i\left(2M^3\tau\right)
    \left(
      n+\frac{b+2Mr}{2M^2}
    \right)^2
  \right]\nonumber\\
  ={}&
  \sum_{r=0}^{M-1}\sum_{n\in\mathbf{Z}}
  \exp\!\left[
    \frac{\pi i\tau}{2M}
    \left(b+2M(r+Mn)\right)^2
  \right]\nonumber\\
  ={}&
  \sum_{s\in\mathbf{Z}}
  \exp\!\left[
    \frac{\pi i\tau}{2M}(b+2Ms)^2
  \right]\nonumber
  \\
  ={}&
    \vartheta\begin{bmatrix}
      \dfrac{b}{2M} \\
      0
  \end{bmatrix}
    \left(0,2M\tau\right)
  ,
  \label{Thetab}
\end{align}
where we have used $s=r+Mn$. Using the properties of the Jacobi theta function, we find that $\Theta_b(\tau)$ satisfies the following relations:
\begin{align}
    \Theta_{b+2M}=\Theta_{b}, \qquad  \Theta_{-b}= \Theta_{b}.
    \label{Jacobisym}
\end{align}
These relations reduce the number of independent $\Theta_b$ from $2M$ to $M+1$. In orbifold models, this implies that some Yukawa couplings allowed by the selection rule may nevertheless vanish identically. Indeed, different combinations of $(i,j,a)$ can give rise to the same $\Theta_b$. If the corresponding orbifold expansion coefficients sum to zero, all contributions proportional to that $\Theta_b$ cancel exactly.

The above analysis has applied to the case with the trivial SS phase $(0,0)$. We now consider the case with nontrivial SS phases, for which the Yukawa coupling takes the Eq. \eqref{eq:yukawa-coupling}. The symmetry properties of the Jacobi theta function are determined by the part excluding the overall factor and the exponential factor. However, these factors cannot be neglected when analyzing cancellations among the corresponding coefficients. For the flux configuration of the form $(M,M,2M)$, the Jacobi theta function can be written as follows,

\begin{align}
&\sum_{m=0}^{2M-1}
\vartheta
\begin{bmatrix}
\dfrac{i-j+(\alpha_1^{(i)}-\alpha_1^{(j)})+Mm}{2M^2}\\0
\end{bmatrix}
\left(M(\alpha_{\tau}^{(j)}-\alpha_{\tau}^{(i)}),
2M^3\tau\right) \notag \\ & \times \delta_{i+j-a+(\alpha_1^{(i)}+\alpha_1^{(j)}-\alpha_1^{(a)}),\,2M\ell -Mm}.
\label{eq:SSphaseJacobi}
\end{align}
The condition due to the Kronecker delta is given by
\begin{align}
    i+j-a+(\alpha_1^{(i)}+\alpha_1^{(j)}-\alpha_1^{(a)})+Mm=0\quad (\mathrm{mod}\ 2M).
\end{align}
Decomposing $m = 2r + \epsilon$ as before, the selection rule becomes
\begin{align}
    i + j - a + (\alpha_1^{(i)} + \alpha_1^{(j)} - \alpha_1^{(a)}) + M\epsilon \equiv 0 \pmod{2M},
\end{align}
where $\alpha_1^{(i)} + \alpha_1^{(j)} - \alpha_1^{(a)} \in \mathbb{Z}$, so that $\epsilon$ is uniquely fixed.
We then define
\begin{align}
    c = i - j + (\alpha_1^{(i)} - \alpha_1^{(j)}) + M\epsilon \pmod{2M},\qquad \beta \equiv \alpha_\tau^{(j)} - \alpha_\tau^{(i)}   . 
\end{align}
Note that $c$ is a half-integer when $\alpha_1^{(i)} \neq \alpha_1^{(j)}$. Then, provided that the coupling satisfies the selection rule, it can be written as
\begin{align}
    \Theta_c^{(\beta)}(\tau) = \sum_{r=0}^{M-1} \vartheta \begin{bmatrix} \dfrac{c + 2Mr}{2M^2} \\ 0 \end{bmatrix}(M\beta, 2M^3\tau) = \vartheta \begin{bmatrix} \dfrac{c}{2M} \\ 0 \end{bmatrix}(\beta, 2M\tau).
\end{align}
Here, by using the definition of the Jacobi theta function, we can write 
\begin{align}
\Theta_{-c}^{(\beta)}(\tau)
&=
\sum_{n\in\mathbb Z}
\exp\left[
\pi i
\left(n-\frac{c}{2M}\right)^2
2M\tau
+
2\pi i
\left(n-\frac{c}{2M}\right)\beta
\right]\nonumber\\
&=\sum_{l\in\mathbb Z}
\exp\left[
\pi i
\left(l+\frac{c}{2M}\right)^2
2M\tau
-
2\pi i
\left(l+\frac{c}{2M}\right)\beta
\right]\nonumber\\
&=\Theta_{c}^{(-\beta)}(\tau),
\end{align}
where we have made the change of summation variable \(n=-l\). Furthermore, we note that the following relation holds for $q\in\mathbf{Z}$:

\begin{align}
\Theta_c^{(\beta-q)}(\tau)&=\vartheta \begin{bmatrix} \dfrac{c}{2M} \\ 0 \end{bmatrix}(\beta-q, 2M\tau)\nonumber\\
&=
\sum_{n\in\mathbb Z}
\exp\left[
\pi i
\left(n+\frac{c}{2M}\right)^2
2M\tau
+
2\pi i
\left(n+\frac{c}{2M}\right)(\beta-q)
\right]\nonumber\\
&=\exp\left(-\frac{\pi icq}{M}\right)\Theta_{c}(\tau).
\label{eq:b-q}
\end{align}
Since $\Theta_c^{(\beta)}(\tau) = \vartheta \begin{bmatrix} c/2M \\ 0 \end{bmatrix}(-\beta, 2M\tau)$ and $-\beta = \beta - 2\beta$, we can apply Eq. \eqref{eq:b-q} with $q = 2\beta$, which is allowed only when $2\beta \in \mathbb{Z}$. We then obtain
\begin{align}
    \Theta_{-c}^{(\beta)}(\tau) = \exp\left(-\frac{2\pi ic\beta}{M}\right)\Theta_c^{(\beta)}(\tau), \quad (2\beta \in \mathbb{Z})  .  
\end{align}
For SS phases taking values in $\{0, 1/2\}$, we have $\beta \in \{0, \pm 1/2\}$ and this condition is always satisfied; for other values of $\beta$, no such proportionality holds in general.
For $\beta=0$, we obtain 
\begin{align}
    \Theta_{-c}^{(0)}= \Theta_{c}^{(0)}.
    \label{}
\end{align}
On the other hand, for $\beta=\pm 
1/2$, it becomes
\begin{align}
    \Theta_{-c}^{(\pm1/2)}= \exp\left(\mp\frac{\pi ic}{M}\right)\Theta_{c}^{(\pm1/2)}.
    \label{}
\end{align}
We can also obtain $\Theta_{c+2M}^{(\beta)} = \Theta_c^{(\beta)}$, which holds for arbitrary $\beta$, as seen by shifting $n \to n - 1$.

The symmetry properties of the Jacobi theta function described above imply that, upon expanding the $T^2/\mathbf{Z}_2$ or $T^2/\mathbf{Z}_4$ wavefunctions in terms of the $T^2$ zero-mode basis, different combinations of $(i,j,a)$ can give rise to identical Jacobi theta functions. If the corresponding coefficients cancel among such contributions, their total contribution vanishes identically.

Specifically, we investigate the $T^2/\mathbf{Z}_2$ orbifold models. Denoting such additional zeros by $C$, the resulting textures are summarized in Table \ref{tab:textureT2Z2-C}. 
In the analysis of section \ref{sec:Textures}, the NNI texture is statistically favored as shown in Table \ref{tab:texture_classification}. 
However, eleven cases leading the NNI textures become the texture 5 in Table \ref{tab:texture_classification} by additional cancellations, as shown in Table \ref{tab:textureT2Z2-C}.
On the other hand, six cases leading to the texture 1 in Table \ref{tab:texture_classification} become the NNI texture by additional cancellations, as shown in Table \ref{tab:textureT2Z2-C}.
Thus, the NNI texture is still statistically favored in magnetized $T^2/\mathbf{Z}_2$ orbifold models.

Note that the above additional cancellation does not occur in the models without SS phases, which correspond to patterns 1 and 2 in Table \ref{tab1}.
Thus, the additional cancellations are important in $T^2/\mathbf{Z}_2$ orbifold models with SS phases. 
The selection rules of $T^2/\mathbf{Z}_2$ orbifold models with SS phases discussed in the previous section are similar to those of $T^2/\mathbf{Z}_2$ orbifold models without SS phases, i.e. $\mathbf{Z}_2$ gauging of $\mathbf{Z}_M$ symmetries.
However, when we take the additional cancellations into account, the selection rules between models with and without SS phases are different from each other.

\begin{table}[H]
    \centering
    \begin{tabular}{|c|c|c|c|c|c|c|}
        \hline
         Pattern & $(M_1,M_2,M_3)$ & \multicolumn{5}{c|}{Yukawa textures} \\
        \hline
         6 & $(5,5,10)$ & 
            $\begin{pmatrix}
                * & 0 & 0 \\
                0 & * & 0 \\
                0 & 0 & *
            \end{pmatrix}$& 
            $\begin{pmatrix}
                0 & * & 0 \\
                * & 0 & * \\
                0 & * & C
            \end{pmatrix}$& 
            $\begin{pmatrix}
                0 & 0 & * \\
                0 & * & * \\
                * & * & 0
            \end{pmatrix}$& 
            $\begin{pmatrix}
                0 & 0 & * \\
                0 & C & * \\
                * & * & 0
            \end{pmatrix}$&
            $\begin{pmatrix}
                 0 & * & 0 \\
                * & 0 & * \\
                0 & * & *
            \end{pmatrix}$  \\
        \hline
         \multirow{2}{*}{8} &\multirow{2}{*}{$(7,7,14)$}&
            $\begin{pmatrix}
                * & 0 & 0 \\
                0 & * & 0 \\
                0 & 0 & *
            \end{pmatrix}$& 
            $\begin{pmatrix}
                0 & * & 0 \\
                * & 0 & * \\
                0 & * & C
            \end{pmatrix}$& 
            $\begin{pmatrix}
                * & 0 & * \\
                0 & 0 & * \\
                * & * & 0
            \end{pmatrix}$& 
            $\begin{pmatrix}
                0 & * & * \\
                * & C & 0 \\
                * & 0 & 0
            \end{pmatrix}$&
            $\begin{pmatrix}
                0 & * & * \\
                * & * & 0 \\
                * & 0 & 0
            \end{pmatrix}$  \\
           & & $\begin{pmatrix}
                C & 0 & * \\
                0 & 0 & * \\
                * & * & 0
            \end{pmatrix}$  &
               $\begin{pmatrix}
                0 & * & * \\
                * & * & 0 \\
                * & 0 & 0
            \end{pmatrix}$ & & & \\
        \hline
        
         \multirow{2}{*}{12} &\multirow{2}{*}{$(6,6,12)$}&
            $\begin{pmatrix}
                * & 0 & 0 \\
                0 & * & 0 \\
                0 & 0 & *
            \end{pmatrix}$& 
            $\begin{pmatrix}
                0 & * & 0 \\
                * & 0 & * \\
                0 & * & C
            \end{pmatrix}$& 
            $\begin{pmatrix}
                0 & * & * \\
                * & 0 & * \\
                * & * & 0
            \end{pmatrix}$& 
            $\begin{pmatrix}
                0 & 0 & *\\
                0 & * & 0 \\
                * & 0 & 0
            \end{pmatrix}$&
            $\begin{pmatrix}
                0 & * & * \\
                * & 0 & * \\
                * & * & 0
            \end{pmatrix}$\\ 
            & & 
            $\begin{pmatrix}
                C & * & 0 \\
                * & 0 & * \\
                0 & * & *
            \end{pmatrix}$ & & & &
            \\
        \hline
        \multirow{2}{*}{13} & \multirow{2}{*}{$(6,6,12)$}&
            $\begin{pmatrix}
                * & 0 & 0 \\
                0 & * & 0 \\
                0 & 0 & *
            \end{pmatrix}$& 
            $\begin{pmatrix}
                * & * & 0 \\
                * & 0 & * \\
                0 & * & C
            \end{pmatrix}$& 
            $\begin{pmatrix}
                0 & * & * \\
                * & 0 & * \\
                * & * & 0
            \end{pmatrix}$& 
            $\begin{pmatrix}
                0 & 0 & * \\
                0 & * & 0 \\
                * & 0 & 0
            \end{pmatrix}$&
            $\begin{pmatrix}
                0 & * & * \\
                * & 0 & * \\
                * & * & 0
            \end{pmatrix}$\\
        & & 
            $\begin{pmatrix}
                C & * & 0 \\
                * & 0 & * \\
                0 & * & *
            \end{pmatrix}$ & & & &\\
            \hline
        14 &$(5,5,10)$&
            $\begin{pmatrix}
                * & * & 0 \\
                * & 0 & * \\
                0 & * & 0
            \end{pmatrix}$& 
            $\begin{pmatrix}
                0 & * & * \\
                * & C & 0 \\
                * & 0 & 0
            \end{pmatrix}$& 
            $\begin{pmatrix}
                0 & * & * \\
                * & * & 0 \\
                * & 0 & 0
            \end{pmatrix}$& 
            $\begin{pmatrix}
                C & * & 0 \\
                * & 0 & * \\
                0 & * & 0
            \end{pmatrix}$&
            $\begin{pmatrix}
                * & 0 & 0 \\
                0 & * & 0 \\
                0 & 0 & *
            \end{pmatrix}$\\ 
        \hline
        \multirow{2}{*}{14} &\multirow{2}{*}{$(6,6,12)$}&
            $\begin{pmatrix}
                * & * & 0 \\
                * & 0 & * \\
                0 & * & C
            \end{pmatrix}$& 
            $\begin{pmatrix}
                0 & * & * \\
                * & 0 & * \\
                * & * & 0
            \end{pmatrix}$& 
            $\begin{pmatrix}
                0 & 0 & * \\
                0 & * & 0 \\
                * & 0 & 0
            \end{pmatrix}$& 
            $\begin{pmatrix}
                0 & * & * \\
                * & 0 & * \\
                * & * & 0
            \end{pmatrix}$&
            $\begin{pmatrix}
                C & * & 0 \\
                * & 0 & * \\
                0 & * & *
            \end{pmatrix}$\\ 
            & & 
            $\begin{pmatrix}
                * & 0 & 0 \\
                0 & * & 0 \\
                0 & 0 & *
            \end{pmatrix}$ & & & &
            \\
        \hline
        \multirow{2}{*}{15} &\multirow{2}{*}{$(6,6,12)$}&
            $\begin{pmatrix}
               * & * & 0 \\
                * & 0 & * \\
                0 & * & C
            \end{pmatrix}$& 
            $\begin{pmatrix}
                0 & * & * \\
                * & 0 & * \\
                * & * & 0
            \end{pmatrix}$& 
            $\begin{pmatrix}
                0 & 0 & * \\
                0 & * & 0 \\
                * & 0 & 0
            \end{pmatrix}$& 
            $\begin{pmatrix}
                0 & * & * \\
                * & 0 & * \\
                * & * & 0
            \end{pmatrix}$&
            $\begin{pmatrix}
                C & * & 0 \\
                * & 0 & * \\
                0 & * & *
            \end{pmatrix}$\\ 
            & & 
            $\begin{pmatrix}
                * & 0 & 0 \\
                0 & * & 0 \\
                0 & 0 & *
            \end{pmatrix}$ &  & & &
            \\
        \hline
        \multirow{2}{*}{15} &\multirow{2}{*}{$(7,7,14)$}&
            $\begin{pmatrix}
                * & * & 0 \\
                * & 0 & * \\
                0 & * & 0
            \end{pmatrix}$& 
            $\begin{pmatrix}
                0 & * & * \\
                * & 0 & 0 \\
                * & 0 & C
            \end{pmatrix}$& 
            $\begin{pmatrix}
                0 & 0 & * \\
                0 & * & * \\
                * & * & 0
            \end{pmatrix}$& 
            $\begin{pmatrix}
                0 & 0 & * \\
                0 & C & * \\
                * & * & 0
            \end{pmatrix}$&
            $\begin{pmatrix}
                0 & * & * \\
                * & 0 & 0 \\
                * & 0 & *
            \end{pmatrix}$\\ 
            & & 
            $\begin{pmatrix}
                C & * & 0 \\
                * & 0 & * \\
                0 & * & 0
            \end{pmatrix}$ & $\begin{pmatrix}
                * & 0 & 0 \\
                0 & * & 0 \\
                0 & 0 & *
            \end{pmatrix}$& &  &
            \\
        \hline
    \end{tabular}
    \caption{Additional cancellations in Yukawa textures on the \(T^2/\mathbf{Z}_2\) orbifold.}
    \label{tab:textureT2Z2-C}
\end{table}

\begin{table}[H]
    \centering
    \begin{tabular}{|c|c|c|c|c|c|c|}
        \hline
         $(M_1,M_2,M_3)$ & $(\gamma_1,\gamma_2,\gamma_3^{*})$ & \multicolumn{5}{c|}{Yukawa textures} \\
        \hline
         (8,8,16) & (1,1,1) & 
            $\begin{pmatrix}
                0 & C & * \\
                C & * & C \\
                * & C & 0
            \end{pmatrix}$& 
            $\begin{pmatrix}
                C & * & C \\
                * & C & * \\
                C & * & C
            \end{pmatrix}$& 
            $\begin{pmatrix}
                C & * & C \\
                * & C & * \\
                C & * & C
            \end{pmatrix}$& 
            $\begin{pmatrix}
                C & C & 0 \\
                C & * & C \\
                0 & C & *
            \end{pmatrix}$&
            $\begin{pmatrix}
                * & C & 0 \\
                C & * & C \\
                0 & C & *
            \end{pmatrix}$  \\
        \hline
         (9,9,18) & (1,1,1) &
            $\begin{pmatrix}
                * & * & * \\
                * & 0 & C \\
                * & C & *
            \end{pmatrix}$& 
            $\begin{pmatrix}
                * & * & * \\
                * & 0 & C \\
                * & C & *
            \end{pmatrix}$& 
            $\begin{pmatrix}
                * & 0 & C \\
                0 & * & C \\
                C & C & *
            \end{pmatrix}$& 
            $\begin{pmatrix}
                * & C & * \\
                C & C & * \\
                * & * & *
            \end{pmatrix}$&
            $\begin{pmatrix}
                * & C & * \\
                C & C & * \\
                * & * & *
            \end{pmatrix}$  \\
        \hline
        (10,10,20) &$(1,-1,-1)$&
            $\begin{pmatrix}
                * & * & * \\
                * & * & * \\
                * & * & C
            \end{pmatrix}$& 
            $\begin{pmatrix}
                * & * & * \\
                * & * & * \\
                * & * & *
            \end{pmatrix}$& 
            $\begin{pmatrix}
                * & * & * \\
                * & * & * \\
                * & * & *
            \end{pmatrix}$& 
            $\begin{pmatrix}
                * & * & * \\
                * & * & * \\
                * & * & *
            \end{pmatrix}$&
            $\begin{pmatrix}
                * & * & * \\
                * & * & * \\
                * & * & *
            \end{pmatrix}$
            \\
        \hline
         \multirow{2}{*}{(12,12,24)} &\multirow{2}{*}{$(i,i,-1)$}&
            $\begin{pmatrix}
                C & C & C \\
                C & C & * \\
                C & * & C
            \end{pmatrix}$& 
            $\begin{pmatrix}
                C & * & * \\
                * & C & * \\
                * & * & *
            \end{pmatrix}$& 
            $\begin{pmatrix}
                C & * & * \\
                * & C & * \\
                * & * & *
            \end{pmatrix}$& 
            $\begin{pmatrix}
                C & * & * \\
                * & C & * \\
                * & * & *
            \end{pmatrix}$&
            $\begin{pmatrix}
                * & 0 & C \\
                0 & * & C \\
                C & C & *
            \end{pmatrix}$\\ 
            & & 
            $\begin{pmatrix}
                * & 0 & C \\
                0 & * & C \\
                C & C & *
            \end{pmatrix}$ & & & &
            \\
        \hline
        (12,12,24) &$(i,-1,i)$&
            $\begin{pmatrix}
                * & * & C \\
                C & C & C \\
                * & * & *
            \end{pmatrix}$& 
            $\begin{pmatrix}
                * & * & * \\
                * & * & * \\
                * & * & *
            \end{pmatrix}$& 
            $\begin{pmatrix}
                * & * & * \\
                * & * & * \\
                * & * & C
            \end{pmatrix}$& 
            $\begin{pmatrix}
                * & * & * \\
                * & * & * \\
                * & * & *
            \end{pmatrix}$&
            $\begin{pmatrix}
                * & * & * \\
                * & * & * \\
                * & * & *
            \end{pmatrix}$\\
        \hline
        \multirow{2}{*}{(14,14,28)} &\multirow{2}{*}{$(i,-i,1)$}&
            $\begin{pmatrix}
                * & * & * \\
                * & * & * \\
                * & * & C
            \end{pmatrix}$& 
            $\begin{pmatrix}
                * & * & * \\
                * & * & * \\
                * & * & C
            \end{pmatrix}$& 
            $\begin{pmatrix}
                * & * & * \\
                * & * & * \\
                * & * & *
            \end{pmatrix}$& 
            $\begin{pmatrix}
                * & * & * \\
                * & * & * \\
                * & * & *
            \end{pmatrix}$&
            $\begin{pmatrix}
                * & * & * \\
                * & * & * \\
                * & * & *
            \end{pmatrix}$\\ 
            & & 
            $\begin{pmatrix}
                * & * & * \\
                * & * & * \\
                * & * & *
            \end{pmatrix}$ & $\begin{pmatrix}
                * & * & * \\
                * & * & * \\
                * & * & *
            \end{pmatrix}$& $\begin{pmatrix}
                * & * & * \\
                * & * & * \\
                * & * & *
            \end{pmatrix}$& &
            \\
        \hline
        \multirow{2}{*}{(15,15,30)} &\multirow{2}{*}{$(-i,-i,-1)$}&
            $\begin{pmatrix}
                * & * & * \\
                * & C & * \\
                * & * & *
            \end{pmatrix}$& 
            $\begin{pmatrix}
                * & * & * \\
                * & C & * \\
                * & * & *
            \end{pmatrix}$& 
            $\begin{pmatrix}
                * & * & * \\
                * & * & * \\
                * & * & *
            \end{pmatrix}$& 
            $\begin{pmatrix}
                * & * & * \\
                * & * & * \\
                * & * & *
            \end{pmatrix}$&
            $\begin{pmatrix}
                * & * & * \\
                * & * & * \\
                * & * & *
            \end{pmatrix}$\\ 
            & & 
            $\begin{pmatrix}
                * & * & * \\
                * & * & * \\
                * & * & *
            \end{pmatrix}$ & $\begin{pmatrix}
                * & * & * \\
                * & * & * \\
                * & * & *
            \end{pmatrix}$& $\begin{pmatrix}
                * & * & * \\
                * & * & * \\
                * & * & *
            \end{pmatrix}$& &
            \\
        \hline
        \multirow{2}{*}{(16,16,32)} &\multirow{2}{*}{$(-i,-i,-1)$}&
            $\begin{pmatrix}
                * & * & * \\
                * & * & * \\
                * & * & *
            \end{pmatrix}$& 
            $\begin{pmatrix}
                * & * & * \\
                * & * & * \\
                * & * & C
            \end{pmatrix}$& 
            $\begin{pmatrix}
                * & * & * \\
                * & * & * \\
                * & * & *
            \end{pmatrix}$& 
            $\begin{pmatrix}
                * & * & * \\
                * & * & * \\
                * & * & C
            \end{pmatrix}$&
            $\begin{pmatrix}
                * & * & * \\
                * & * & * \\
                * & * & C
            \end{pmatrix}$\\ 
            & & 
            $\begin{pmatrix}
                * & * & * \\
                * & * & * \\
                * & * & C
            \end{pmatrix}$ & $\begin{pmatrix}
                * & * & C \\
                * & * & C \\
                C & C & *
            \end{pmatrix}$& $\begin{pmatrix}
                * & * & * \\
                * & * & * \\
                * & * & *
            \end{pmatrix}$& &
            \\
        \hline
    \end{tabular}
    \caption{Additional cancellations in Yukawa textures on the \(T^2/\mathbf{Z}_4\) orbifold.}
    \label{tab:texture00-C}
\end{table}

We can also investigate the $T^2/\mathbf{Z}_4$ orbifold models. Additional vanishing entries also appear in Yukawa textures for which all entries are allowed by the selection rule. The resulting textures are summarized in Table \ref{tab:texture00-C}. 
Such additional cancellations do not occur except  in the cases shown in Table \ref{tab:texture00-C}.
In several entries, additional cancellations appear.
Note that the additional cancellations occur even in $T^2/\mathbf{Z}_4$ models without SS phases, although such cancellations do not occur in $T^2/\mathbf{Z}_2$ orbifold models without SS phases.
These textures are non-trivial, and some of them cannot be realized by group-theoretical symmetries.
To summarize only rank-3 textures, we can derive the following six textures:
\begin{align}
    \begin{pmatrix}
        * & * & * \\
        * & * & * \\
        * & * & * \\
    \end{pmatrix}, \qquad 
     \begin{pmatrix}
        0 & * & * \\
        * & * & * \\
        * & * & * \\
    \end{pmatrix}, \qquad 
     \begin{pmatrix}
        0 & * & * \\
        * & 0 & * \\
        * & * & * \\ 
    \end{pmatrix}, \qquad 
      \begin{pmatrix}
        * & 0 & * \\
        0 & 0 & * \\
        * & * & * \\ 
    \end{pmatrix}, \qquad 
     \begin{pmatrix}
        * & 0 & 0 \\
        0 & * & * \\
        0 & * & * \\ 
    \end{pmatrix}, \qquad 
     \begin{pmatrix}
        * & 0 & 0 \\
        0 & * & 0 \\
        0 & 0 & * \\ 
    \end{pmatrix}, \qquad 
\end{align}
in magnetized $T^2/\mathbf{Z}_4$ orbifold models, where we just denote $C$ by $0$.

%%%%%%%%%%%%%%%%%%%%%%%%%%%%%%%%%%%%%%%%%%%%%%%%%%%%%%%%%%%%%%%%%
\section{Conclusion}
\label{sec:conclusion}

We have studied systematically Yukawa textures, which can be derived from magnetized $T^2/\mathbf{Z}_2$ and $T^2/\mathbf{Z}_4$ orbifold models with various magnetic fluxes, SS phases, and $\mathbf{Z}_2$ and $\mathbf{Z}_4$ eigenvalues.
In magnetized $T^2$ models, there appear the exact $\mathbf{Z}_M$ symmetries, which control Yukawa couplings and other couplings.
These $\mathbf{Z}_M$ symmetries are violated on orbifolds, but some coupling selection rules remain.
In magnetized $T^2/\mathbf{Z}_2$ orbifold models without SS phases, the coupling selection rules are controlled by $\mathbf{Z}_2$ gauging of $\mathbf{Z}_M$ symmetries, as shown in Refs.~\cite{Kobayashi:2024yqq,Kobayashi:2024cvp}.
Magnetized $T^2/\mathbf{Z}_2$ orbifold models with SS phases have similar selection rules.
Then, nontrivial textures are derived in magnetized $T^2/\mathbf{Z}_2$ orbifold models with and without SS phases.
It is remarkable that the NNI form and the diagonal form are statistically favorable.
On top of that, if the flavor assignments are the same between left-handed and right-handed fermions, the diagonal form is statistically favored.
On the other hand, if the flavor assignments are different between left-handed and right-handed fermions, the NNI form is favored.
Also, in certain magnetized $T^2/\mathbf{Z}_4$ orbifold models, nontrivial selection rules, which originate from the $\mathbf{Z}_M$ symmetries, remain, and nontrivial textures are realized.

Furthermore, when the flux configuration corresponds to $(M,M,2M)$, additional cancellations appear and they lead to nontrivial textures.
When we take these additional cancellations into account, the selection rules between  $T^2/\mathbf{Z}_2$ orbifold models with and without SS phases are different from each other at least for three-generation models.
In magnetized $T^2/\mathbf{Z}_2$ orbifold models, the NNI texture is statistically favored even when we take such additional cancellations into account.
The textures have richer structure in $T^2/\mathbf{Z}_4$ orbifold models by such additional cancellations.

We have shown that coupling selection rules and additional cancellations in magnetized $T^2/\mathbf{Z}_2$ and $T^2/\mathbf{Z}_4$ orbifold models are interesting.
We have applied them to derive nontrivial textures.
It is interesting to apply them to other physical aspects such as neutrino physics,   dark matter physics, and SMEFT operators.
Also numerical studies on Yukawa matrices are important.
We would study them elsewhere.

Moreover, localized magnetic fluxes at orbifold fixed points induce additional localized modes \cite{Lee:2003mc,Buchmuller:2015eya,Buchmuller:2018lkz,Abe:2020vmv,Kobayashi:2022tti}.
It would be important to study selection rules including these localized modes and bulk modes.
We leave this for future work.

\acknowledgments

This work was supported by JSPS KAKENHI Grant Numbers JP23K03375 (T.K.), and JST SPRING, Grant Number JPMJSP2119 (H.U.).

\bibliography{references}{}

@article{Weinberg:1977hb,
    author = "Weinberg, Steven",
    title = "{The Problem of Mass}",
    reportNumber = "HUTP-77-A057",
    doi = "10.1111/j.2164-0947.1977.tb02958.x",
    journal = "Trans. New York Acad. Sci.",
    volume = "38",
    pages = "185--201",
    year = "1977"
}

@article{Fritzsch:1977vd,
    author = "Fritzsch, H.",
    title = "{Weak Interaction Mixing in the Six - Quark Theory}",
    reportNumber = "CERN-TH-2433",
    doi = "10.1016/0370-2693(78)90524-5",
    journal = "Phys. Lett. B",
    volume = "73",
    pages = "317--322",
    year = "1978"
}

@article{Kobayashi:2024cvp,
    author = "Kobayashi, Tatsuo and Otsuka, Hajime and Tanimoto, Morimitsu",
    title = "{Yukawa textures from non-invertible symmetries}",
    eprint = "2409.05270",
    archivePrefix = "arXiv",
    primaryClass = "hep-ph",
    reportNumber = "EPHOU-24-013, KYUSHU-HET-295",
    doi = "10.1007/JHEP12(2024)117",
    journal = "JHEP",
    volume = "12",
    pages = "117",
    year = "2024"
}

@article{Kobayashi:2025znw,
    author = "Kobayashi, Tatsuo and Nishioka, Yume and Otsuka, Hajime and Tanimoto, Morimitsu",
    title = "{More about quark Yukawa textures from selection rules without group actions}",
    eprint = "2503.09966",
    archivePrefix = "arXiv",
    primaryClass = "hep-ph",
    reportNumber = "EPHOU-25-003, KYUSHU-HET-312",
    doi = "10.1007/JHEP05(2025)177",
    journal = "JHEP",
    volume = "05",
    pages = "177",
    year = "2025"
}

@article{Suzuki:2025oov,
    author = "Suzuki, Motoo and Xu, Ling-Xiao",
    title = "{Phenomenological implications of a class of non-invertible selection rules}",
    eprint = "2503.19964",
    archivePrefix = "arXiv",
    primaryClass = "hep-ph",
    month = "3",
    year = "2025"
}

@article{Liang:2025dkm,
    author = "Liang, Qiuyue and Yanagida, Tsutomu T.",
    title = "{Non-invertible symmetry as an axion-less solution to the strong CP problem}",
    eprint = "2505.05142",
    archivePrefix = "arXiv",
    primaryClass = "hep-ph",
    doi = "10.1016/j.physletb.2025.139706",
    journal = "Phys. Lett. B",
    volume = "868",
    pages = "139706",
    year = "2025"
}

@article{Abe:2008fi,
    author = "Abe, Hiroyuki and Kobayashi, Tatsuo and Ohki, Hiroshi",
    title = "{Magnetized orbifold models}",
    eprint = "0806.4748",
    archivePrefix = "arXiv",
    primaryClass = "hep-th",
    reportNumber = "YITP-08-48, KUNS-2145",
    doi = "10.1088/1126-6708/2008/09/043",
    journal = "JHEP",
    volume = "09",
    pages = "043",
    year = "2008"
}

@article{Cremades:2004wa,
    author = "Cremades, D. and Ibanez, L. E. and Marchesano, F.",
    title = "{Computing Yukawa couplings from magnetized extra dimensions}",
    eprint = "hep-th/0404229",
    archivePrefix = "arXiv",
    reportNumber = "FTUAM-04-7, IFT-UAM-CSIC-04-15, MAD-TH-04-4",
    doi = "10.1088/1126-6708/2004/05/079",
    journal = "JHEP",
    volume = "05",
    pages = "079",
    year = "2004"
}

@article{Abe:2009vi,
    author = "Abe, Hiroyuki and Choi, Kang-Sin and Kobayashi, Tatsuo and Ohki, Hiroshi",
    title = "{Non-Abelian Discrete Flavor Symmetries from Magnetized/Intersecting Brane Models}",
    eprint = "0904.2631",
    archivePrefix = "arXiv",
    primaryClass = "hep-ph",
    reportNumber = "WU-HEP-09-01, KUNS-2202, YITP-09-29",
    doi = "10.1016/j.nuclphysb.2009.05.024",
    journal = "Nucl. Phys. B",
    volume = "820",
    pages = "317--333",
    year = "2009"
}

@article{Berasaluce-Gonzalez:2012abm,
    author = "Berasaluce-Gonzalez, M. and Camara, P. G. and Marchesano, F. and Regalado, D. and Uranga, A. M.",
    title = "{Non-Abelian discrete gauge symmetries in 4d string models}",
    eprint = "1206.2383",
    archivePrefix = "arXiv",
    primaryClass = "hep-th",
    reportNumber = "IFT-UAM-CSIC-12-53",
    doi = "10.1007/JHEP09(2012)059",
    journal = "JHEP",
    volume = "09",
    pages = "059",
    year = "2012"
}

@article{Marchesano:2013ega,
    author = "Marchesano, Fernando and Regalado, Diego and Vazquez-Mercado, Liliana",
    title = "{Discrete flavor symmetries in D-brane models}",
    eprint = "1306.1284",
    archivePrefix = "arXiv",
    primaryClass = "hep-th",
    reportNumber = "IFT-UAM-CSIC-13-065",
    doi = "10.1007/JHEP09(2013)028",
    journal = "JHEP",
    volume = "09",
    pages = "028",
    year = "2013"
}

@article{Abe:2013bca,
    author = "Abe, Tomo-Hiro and Fujimoto, Yukihiro and Kobayashi, Tatsuo and Miura, Takashi and Nishiwaki, Kenji and Sakamoto, Makoto",
    title = "{$Z_N$ twisted orbifold models with magnetic flux}",
    eprint = "1309.4925",
    archivePrefix = "arXiv",
    primaryClass = "hep-th",
    reportNumber = "KOBE-TH-13-07, KUNS-2463, HRI-P-13-09-001, RECAPP-HRI-2013-019",
    doi = "10.1007/JHEP01(2014)065",
    journal = "JHEP",
    volume = "01",
    pages = "065",
    year = "2014"
}

@article{Abe:2014noa,
    author = "Abe, Tomo-hiro and Fujimoto, Yukihiro and Kobayashi, Tatsuo and Miura, Takashi and Nishiwaki, Kenji and Sakamoto, Makoto",
    title = "{Operator analysis of physical states on magnetized $T^{2}/Z_{N}$ orbifolds}",
    eprint = "1409.5421",
    archivePrefix = "arXiv",
    primaryClass = "hep-th",
    reportNumber = "EHOU-14-003, HRI-P-14-02-001, KOBE-TH-14-02, KUNS-2485, OU-HET-827, RECAPP-HRI-2014-004",
    doi = "10.1016/j.nuclphysb.2014.11.022",
    journal = "Nucl. Phys. B",
    volume = "890",
    pages = "442--480",
    year = "2014"
}

@article{Kobayashi:2017dyu,
    author = "Kobayashi, Tatsuo and Nagamoto, Satoshi",
    title = "{Zero-modes on orbifolds : magnetized orbifold models by modular transformation}",
    eprint = "1709.09784",
    archivePrefix = "arXiv",
    primaryClass = "hep-th",
    reportNumber = "EPHOU-17-013",
    doi = "10.1103/PhysRevD.96.096011",
    journal = "Phys. Rev. D",
    volume = "96",
    number = "9",
    pages = "096011",
    year = "2017"
}

@article{Hamidi:1986vh,
    author = "Hamidi, Shahram and Vafa, Cumrun",
    title = "{Interactions on Orbifolds}",
    reportNumber = "HUTP-86-A041, CALT-68-1349",
    doi = "10.1016/0550-3213(87)90006-X",
    journal = "Nucl. Phys. B",
    volume = "279",
    pages = "465--513",
    year = "1987"
}

@article{Dixon:1986qv,
    author = "Dixon, Lance J. and Friedan, Daniel and Martinec, Emil J. and Shenker, Stephen H.",
    title = "{The Conformal Field Theory of Orbifolds}",
    reportNumber = "EFI-86-42-CHICAGO",
    doi = "10.1016/0550-3213(87)90676-6",
    journal = "Nucl. Phys. B",
    volume = "282",
    pages = "13--73",
    year = "1987"
}

@article{Kaidi:2024wio,
    author = "Kaidi, Justin and Tachikawa, Yuji and Zhang, Hao Y.",
    title = "{On a class of selection rules without group actions in field theory and string theory}",
    eprint = "2402.00105",
    archivePrefix = "arXiv",
    primaryClass = "hep-th",
    doi = "10.21468/SciPostPhys.17.6.169",
    journal = "SciPost Phys.",
    volume = "17",
    number = "6",
    pages = "169",
    year = "2024"
}

@article{Branco:1988iq,
    author = "Branco, G. C. and Lavoura, L. and Mota, Fatima",
    title = "{Nearest Neighbor Interactions and the Physical Content of Fritzsch Mass Matrices}",
    reportNumber = "IFM-27/88",
    doi = "10.1103/PhysRevD.39.3443",
    journal = "Phys. Rev. D",
    volume = "39",
    pages = "3443",
    year = "1989"
}

@article{Branco:1994jx,
    author = "Branco, G. C. and Silva-Marcos, J. I.",
    title = "{NonHermitian Yukawa couplings?}",
    reportNumber = "CERN-TH-7229-94",
    doi = "10.1016/0370-2693(94)91069-3",
    journal = "Phys. Lett. B",
    volume = "331",
    pages = "390--394",
    year = "1994"
}

@article{Branco:1999nb,
    author = "Branco, G. C. and Emmanuel-Costa, D. and Gonzalez Felipe, R.",
    title = "{Texture zeros and weak basis transformations}",
    eprint = "hep-ph/9911418",
    archivePrefix = "arXiv",
    reportNumber = "FISIST-19-99-CFIF",
    doi = "10.1016/S0370-2693(00)00193-3",
    journal = "Phys. Lett. B",
    volume = "477",
    pages = "147--155",
    year = "2000"
}

@article{Kobayashi:2024yqq,
    author = "Kobayashi, Tatsuo and Otsuka, Hajime",
    title = "{Non-invertible flavor symmetries in magnetized extra dimensions}",
    eprint = "2408.13984",
    archivePrefix = "arXiv",
    primaryClass = "hep-th",
    reportNumber = "EPHOU-24-010, KYUSHU-HET-292",
    doi = "10.1007/JHEP11(2024)120",
    journal = "JHEP",
    volume = "11",
    pages = "120",
    year = "2024"
}

@article{Fritzsch:1979zq,
    author = "Fritzsch, Harald",
    title = "{Quark Masses and Flavor Mixing}",
    reportNumber = "CERN-TH-2640",
    doi = "10.1016/0550-3213(79)90362-6",
    journal = "Nucl. Phys. B",
    volume = "155",
    pages = "189--207",
    year = "1979"
}

@article{Ramond:1993kv,
    author = "Ramond, Pierre and Roberts, R. G. and Ross, Graham G.",
    title = "{Stitching the Yukawa quilt}",
    eprint = "hep-ph/9303320",
    archivePrefix = "arXiv",
    reportNumber = "UFIFT-HEP-93-06, RAL-93-010",
    doi = "10.1016/0550-3213(93)90159-M",
    journal = "Nucl. Phys. B",
    volume = "406",
    pages = "19--42",
    year = "1993"
}

@article{Fritzsch:2002ga,
    author = "Fritzsch, Harald and Xing, Zhi-zhong",
    title = "{Four zero texture of Hermitian quark mass matrices and current experimental tests}",
    eprint = "hep-ph/0212195",
    archivePrefix = "arXiv",
    doi = "10.1016/S0370-2693(03)00048-0",
    journal = "Phys. Lett. B",
    volume = "555",
    pages = "63--70",
    year = "2003"
}

@article{Xing:2015sva,
    author = "Xing, Zhi-zhong and Zhao, Zhen-hua",
    title = "{On the four-zero texture of quark mass matrices and its stability}",
    eprint = "1501.06346",
    archivePrefix = "arXiv",
    primaryClass = "hep-ph",
    doi = "10.1016/j.nuclphysb.2015.05.027",
    journal = "Nucl. Phys. B",
    volume = "897",
    pages = "302--325",
    year = "2015"
}

@article{Frampton:2002yf,
    author = "Frampton, Paul H. and Glashow, Sheldon L. and Marfatia, Danny",
    title = "{Zeroes of the neutrino mass matrix}",
    eprint = "hep-ph/0201008",
    archivePrefix = "arXiv",
    reportNumber = "IFP-805-UNC",
    doi = "10.1016/S0370-2693(02)01817-8",
    journal = "Phys. Lett. B",
    volume = "536",
    pages = "79--82",
    year = "2002"
}

@article{Kageyama:2002zw,
    author = "Kageyama, Atsushi and Kaneko, Satoru and Shimoyama, Noriyuki and Tanimoto, Morimitsu",
    title = "{Seesaw realization of the texture zeros in the neutrino mass matrix}",
    eprint = "hep-ph/0204291",
    archivePrefix = "arXiv",
    reportNumber = "NIIG-DP-02-3",
    doi = "10.1016/S0370-2693(02)01964-0",
    journal = "Phys. Lett. B",
    volume = "538",
    pages = "96--106",
    year = "2002"
}

@article{Funakoshi:2024uvy,
    author = "Funakoshi, Shuta and Kobayashi, Tatsuo and Otsuka, Hajime",
    title = "{Quantum aspects of non-invertible flavor symmetries in intersecting/magnetized D-brane models}",
    eprint = "2412.12524",
    archivePrefix = "arXiv",
    primaryClass = "hep-th",
    reportNumber = "EPHOU-24-017, KYUSHU-HET-304",
    doi = "10.1007/JHEP04(2025)183",
    journal = "JHEP",
    volume = "04",
    pages = "183",
    year = "2025"
}

@article{Dixon:1986jc,
    author = "Dixon, Lance J. and Harvey, Jeffrey A. and Vafa, C. and Witten, Edward",
    title = "{Strings on Orbifolds. 2.}",
    reportNumber = "PRINT-86-0246 (PRINCETON)",
    doi = "10.1016/0550-3213(86)90287-7",
    journal = "Nucl. Phys. B",
    volume = "274",
    pages = "285--314",
    year = "1986"
}

@article{Kobayashi:1990mc,
    author = "Kobayashi, Tatsuo and Ohtsubo, Noriyasu",
    title = "{Yukawa Coupling Condition of $Z(N$) Orbifold Models}",
    reportNumber = "DPKU-9002",
    doi = "10.1016/0370-2693(90)90671-R",
    journal = "Phys. Lett. B",
    volume = "245",
    pages = "441--446",
    year = "1990"
}

@article{Kobayashi:1991rp,
    author = "Kobayashi, Tatsuo and Ohtsubo, Noriyasu",
    title = "{Geometrical aspects of Z(N) orbifold phenomenology}",
    reportNumber = "DPKU-9103",
    doi = "10.1142/S0217751X94000054",
    journal = "Int. J. Mod. Phys. A",
    volume = "9",
    pages = "87--126",
    year = "1994"
}

@article{Kobayashi:1995py,
    author = "Kobayashi, Tatsuo",
    title = "{Selection rules for nonrenormalizable couplings in superstring theories}",
    eprint = "hep-ph/9504371",
    archivePrefix = "arXiv",
    reportNumber = "LMU-TPW-95-5",
    doi = "10.1016/0370-2693(95)00643-Y",
    journal = "Phys. Lett. B",
    volume = "354",
    pages = "264--270",
    year = "1995"
}

@article{Kobayashi:2025ocp,
    author = "Kobayashi, Tatsuo and Nishida, Ryusei and Otsuka, Hajime",
    title = "{Non-Invertible Selection Rules on Heterotic Non-Abelian Orbifolds}",
    eprint = "2509.10019",
    archivePrefix = "arXiv",
    primaryClass = "hep-th",
    reportNumber = "EPHOU-25-017, KYUSHU-HET-339",
    month = "9",
    year = "2025"
}

@article{Hernandez-Segura:2026yfx,
    author = "Hernandez-Segura, Miguel and Liu, Xiang-Gan and Perez-Martinez, Ricardo and Ramos-Sanchez, Saul",
    title = "{The Flavor of Non-Abelian Orbifolds}",
    eprint = "2609.30241",
    archivePrefix = "arXiv",
    primaryClass = "hep-th",
    month = "9",
    year = "2026"
}

@article{Bachas:1995ik,
    author = "Bachas, C.",
    title = "{A Way to break supersymmetry}",
    eprint = "hep-th/9503030",
    archivePrefix = "arXiv",
    reportNumber = "CPTH-R349-0395, CPTH-RR349-0395, CPTH-RR.349.0395",
    month = "3",
    year = "1995"
}

@article{Blumenhagen:2000wh,
    author = "Blumenhagen, Ralph and Goerlich, Lars and Kors, Boris and Lust, Dieter",
    title = "{Noncommutative compactifications of type I strings on tori with magnetic background flux}",
    eprint = "hep-th/0007024",
    archivePrefix = "arXiv",
    reportNumber = "HUB-EP-00-27",
    doi = "10.1088/1126-6708/2000/10/006",
    journal = "JHEP",
    volume = "10",
    pages = "006",
    year = "2000"
}

@article{Angelantonj:2000hi,
    author = "Angelantonj, C. and Antoniadis, Ignatios and Dudas, E. and Sagnotti, A.",
    title = "{Type I strings on magnetized orbifolds and brane transmutation}",
    eprint = "hep-th/0007090",
    archivePrefix = "arXiv",
    reportNumber = "LPTENS-00-27, CPTH-S074-0600, LPT-ORSAY-00-61, ROM2F-2000-23, CERN-TH-2000-183",
    doi = "10.1016/S0370-2693(00)00907-2",
    journal = "Phys. Lett. B",
    volume = "489",
    pages = "223--232",
    year = "2000"
}

@article{Blumenhagen:2000ea,
    author = "Blumenhagen, Ralph and Kors, Boris and Lust, Dieter",
    title = "{Type I strings with F flux and B flux}",
    eprint = "hep-th/0012156",
    archivePrefix = "arXiv",
    reportNumber = "HU-EP-00-61",
    doi = "10.1088/1126-6708/2001/02/030",
    journal = "JHEP",
    volume = "02",
    pages = "030",
    year = "2001"
}

@article{Dong:2025jra,
    author = "Dong, Jun and Jeric, Tim and Kobayashi, Tatsuo and Nishida, Ryusei and Otsuka, Hajime",
    title = "{Discrete gauging and noninvertible selection rules}",
    eprint = "2507.02375",
    archivePrefix = "arXiv",
    primaryClass = "hep-th",
    doi = "10.1103/nsvv-l2dy",
    journal = "Phys. Rev. D",
    volume = "113",
    number = "5",
    pages = "056028",
    year = "2026"
}

@article{Kobayashi:2025ldi,
    author = "Kobayashi, Tatsuo and Otsuka, Hajime and Tanimoto, Morimitsu and Uchida, Haruki",
    title = "{Lepton mass textures from non-invertible multiplication rules}",
    eprint = "2505.07262",
    archivePrefix = "arXiv",
    primaryClass = "hep-ph",
    reportNumber = "EPHOU-25-007, KYUSHU-HET-321",
    doi = "10.1007/JHEP08(2025)189",
    journal = "JHEP",
    volume = "08",
    pages = "189",
    year = "2025"
}

@article{Jiang:2025psz,
    author = "Jiang, Zheng and Qu, Bu-Yao and Ding, Gui-Jun",
    title = "{Texture-zeros in minimal seesaw from noninvertible symmetry fusion rules}",
    eprint = "2510.07236",
    archivePrefix = "arXiv",
    primaryClass = "hep-ph",
    doi = "10.1103/d29s-cw34",
    journal = "Phys. Rev. D",
    volume = "112",
    number = "11",
    pages = "115029",
    year = "2025"
}

@article{Qu:2026omn,
    author = "Qu, Bu-Yao and Jiang, Zheng and Ding, Gui-Jun",
    title = "{Two-zero textures of the Majorana neutrino mass matrix from $\mathbb{Z}_3$ gauging of $\mathbb{Z}_N$ non-invertible symmetry}",
    eprint = "2602.24214",
    archivePrefix = "arXiv",
    primaryClass = "hep-ph",
    month = "2",
    year = "2026"
}

@article{Chen:2026mvi,
    author = "Chen, Zi-Qiang and Jiang, Wen-Hao and Zhou, Ye-Ling",
    title = "{Universal two-zero texture in SO(10): implications of JUNO and realization from non-invertible symmetries}",
    eprint = "2606.24571",
    archivePrefix = "arXiv",
    primaryClass = "hep-ph",
    month = "6",
    year = "2026"
}

@article{Kobayashi:2025thd,
    author = "Kobayashi, Tatsuo and Otsuka, Hajime and Yanagida, Tsutomu T.",
    title = "{Noninvertible symmetry as a solution to the strong CP problem in a GUT-inspired standard model}",
    eprint = "2508.12287",
    archivePrefix = "arXiv",
    primaryClass = "hep-ph",
    reportNumber = "EPHOU-25-014, KYUSHU-HET-334",
    doi = "10.1103/vbd5-5cp3",
    journal = "Phys. Rev. D",
    volume = "113",
    number = "5",
    pages = "055016",
    year = "2026"
}

@article{Kobayashi:2025rpx,
    author = "Kobayashi, Tatsuo and Otsuka, Hajime and Tanimoto, Morimitsu and Yanagida, Tsutomu T.",
    title = "{GUT-motivated noninvertible symmetry as a solution to the strong CP problem and the neutrino CP-violating phase}",
    eprint = "2510.01680",
    archivePrefix = "arXiv",
    primaryClass = "hep-ph",
    reportNumber = "EPHOU-25-018, KYUSHU-HET-340",
    doi = "10.1103/f7bw-qlgs",
    journal = "Phys. Rev. D",
    volume = "113",
    number = "9",
    pages = "095034",
    year = "2026"
}

@article{Kobayashi:2025lar,
    author = "Kobayashi, Tatsuo and Mita, Hironobu and Otsuka, Hajime and Sakuma, Riku",
    title = "{Matter Symmetries in Supersymmetric Standard Models from Noninvertible Selection Rules}",
    eprint = "2506.10241",
    archivePrefix = "arXiv",
    primaryClass = "hep-ph",
    reportNumber = "EPHOU-25-009, KYUSHU-HET-325",
    doi = "10.1093/ptep/ptag016",
    journal = "PTEP",
    volume = "2026",
    number = "3",
    pages = "033B01",
    year = "2026"
}

@article{Kobayashi:2025wty,
    author = "Kobayashi, Tatsuo and Otsuka, Hajime",
    title = "{Generalized CP from non-invertible selection rules}",
    eprint = "2512.16376",
    archivePrefix = "arXiv",
    primaryClass = "hep-ph",
    reportNumber = "EPHOU-25-019,KYUSHU-HET-345",
    month = "12",
    year = "2025"
}

@article{Nomura:2026hcu,
    author = "Nomura, Takaaki and Okada, Hiroshi",
    title = "{A theoretical account of tiny multi-Higgs vacuum expectation values from non-invertible symmetry}",
    eprint = "2604.27612",
    archivePrefix = "arXiv",
    primaryClass = "hep-ph",
    month = "4",
    year = "2026"
}

@article{Suzuki:2025bxg,
    author = "Suzuki, Motoo and Xu, Ling-Xiao and Zhang, Hao Y.",
    title = "{Spurion Analysis for Non-Invertible Selection Rules from Near-Group Fusions}",
    eprint = "2508.14970",
    archivePrefix = "arXiv",
    primaryClass = "hep-ph",
    month = "8",
    year = "2025"
}

@article{Suzuki:2025kxz,
    author = "Suzuki, Motoo and Xu, Ling-Xiao",
    title = "{Spurion analysis of {\ensuremath{\mathbb{Z}}}$_{M}$/{\ensuremath{\mathbb{Z}}}$_{2}$ non-invertible selection rules: low-order versus all-order zeros}",
    eprint = "2510.18972",
    archivePrefix = "arXiv",
    primaryClass = "hep-ph",
    doi = "10.1007/JHEP02(2026)227",
    journal = "JHEP",
    volume = "02",
    pages = "227",
    year = "2026"
}

@article{Xu:2026nwh,
    author = "Xu, Ling-Xiao",
    title = "{A General Prescription for Spurion Analysis of Non-Invertible Selection Rules}",
    eprint = "2604.09345",
    archivePrefix = "arXiv",
    primaryClass = "hep-ph",
    month = "4",
    year = "2026"
}

@article{Kobayashi:2025cwx,
    author = "Kobayashi, Tatsuo and Okada, Hiroshi and Otsuka, Hajime",
    title = "{Radiative neutrino mass models from non-invertible selection rules}",
    eprint = "2505.14878",
    archivePrefix = "arXiv",
    primaryClass = "hep-ph",
    reportNumber = "EPHOU-25-008, KYUSHU-HET-324",
    doi = "10.1007/JHEP12(2025)111",
    journal = "JHEP",
    volume = "12",
    pages = "111",
    year = "2025"
}

@article{Nomura:2025sod,
    author = "Nomura, Takaaki and Okada, Hiroshi",
    title = "{Radiative lepton seesaw model in a non-invertible fusion rule and gauged $B-L$ symmetry}",
    eprint = "2506.16706",
    archivePrefix = "arXiv",
    primaryClass = "hep-ph",
    month = "6",
    year = "2025"
}

@article{Chen:2025awz,
    author = "Chen, Jingqian and Geng, Chao-Qiang and Okada, Hiroshi and Wu, Jia-Jun",
    title = "{A radiative lepton model in a non-invertible fusion rule}",
    eprint = "2507.11951",
    archivePrefix = "arXiv",
    primaryClass = "hep-ph",
    doi = "10.1016/j.nuclphysb.2026.117391",
    journal = "Nucl. Phys. B",
    volume = "1025",
    pages = "117391",
    year = "2026"
}

@article{Okada:2025kfm,
    author = "Okada, Hiroshi and Shigekami, Yoshihiro",
    title = "{Three-loop induced neutrino mass model in a non-invertible symmetry}",
    eprint = "2507.16198",
    archivePrefix = "arXiv",
    primaryClass = "hep-ph",
    month = "7",
    year = "2025"
}

@article{Jangid:2025krp,
    author = "Jangid, Shilpa and Okada, Hiroshi",
    title = "{A natural realization of inverse seesaw model in a non-invertible selection rule}",
    eprint = "2508.16174",
    archivePrefix = "arXiv",
    primaryClass = "hep-ph",
    month = "8",
    year = "2025"
}

@article{Nomura:2025tvz,
    author = "Nomura, Takaaki and Okada, Hiroshi and Shigekami, Yoshihiro",
    title = "{Radiative lepton model in a non-invertible fusion rule}",
    eprint = "2510.17156",
    archivePrefix = "arXiv",
    primaryClass = "hep-ph",
    month = "10",
    year = "2025"
}

@article{Okada:2026bpp,
    author = "Okada, Hiroshi and Otsuka, Hajime",
    title = "{Dynamical CP Violation from Non-Invertible Selection Rules}",
    eprint = "2604.04423",
    archivePrefix = "arXiv",
    primaryClass = "hep-ph",
    reportNumber = "KYUSHU-HET-353",
    month = "4",
    year = "2026"
}

@article{Kawamura:2026zae,
    author = "Kawamura, Junichiro",
    title = "{Taming lepton portal dark matter by a non-invertible selection rule}",
    eprint = "2608.29051",
    archivePrefix = "arXiv",
    primaryClass = "hep-ph",
    month = "8",
    year = "2026"
}

@article{Hamada:2026sui,
    author = "Hamada, Yu",
    title = "{A non-invertible Fibonacci selection rule for two-Higgs-doublet models}",
    eprint = "2610.06014",
    archivePrefix = "arXiv",
    primaryClass = "hep-ph",
    reportNumber = "OU-HET-1326",
    month = "10",
    year = "2026"
}

@article{Kobayashi:2026msq,
    author = "Kobayashi, Tatsuo and Otsuka, Hajime and Tanimoto, Morimitsu and Yamamoto, Kei",
    title = "{Textures of dimension-six operators in the SMEFT with non-invertible selection rules}",
    eprint = "2608.21707",
    archivePrefix = "arXiv",
    primaryClass = "hep-ph",
    reportNumber = "EPHOU-26-09, KYUSHU-HET-369",
    month = "8",
    year = "2026"
}

@article{Lee:2003mc,
    author = "Lee, Hyun Min and Nilles, Hans Peter and Zucker, Max",
    title = "{Spontaneous localization of bulk fields: The Six-dimensional case}",
    eprint = "hep-th/0309195",
    archivePrefix = "arXiv",
    doi = "10.1016/j.nuclphysb.2003.12.031",
    journal = "Nucl. Phys. B",
    volume = "680",
    pages = "177--198",
    year = "2004"
}

@article{Buchmuller:2015eya,
    author = "Buchmuller, Wilfried and Dierigl, Markus and Ruehle, Fabian and Schweizer, Julian",
    title = "{Chiral fermions and anomaly cancellation on orbifolds with Wilson lines and flux}",
    eprint = "1506.05771",
    archivePrefix = "arXiv",
    primaryClass = "hep-th",
    reportNumber = "DESY-15-068",
    doi = "10.1103/PhysRevD.92.105031",
    journal = "Phys. Rev. D",
    volume = "92",
    number = "10",
    pages = "105031",
    year = "2015"
}

@article{Buchmuller:2018lkz,
    author = "Buchmuller, Wilfried and Dierigl, Markus and Tatsuta, Yoshiyuki",
    title = "{Magnetized orbifolds and localized flux}",
    eprint = "1810.06362",
    archivePrefix = "arXiv",
    primaryClass = "hep-th",
    reportNumber = "DESY 18-167, DESY-18-167",
    doi = "10.1016/j.aop.2018.12.006",
    journal = "Annals Phys.",
    volume = "401",
    pages = "91--115",
    year = "2019"
}

@article{Abe:2020vmv,
    author = "Abe, Hiroyuki and Kobayashi, Tatsuo and Uemura, Shohei and Yamamoto, Junji",
    title = "{Loop Fayet-Iliopoulos terms in $T^2/Z_2$ models: Instability and moduli stabilization}",
    eprint = "2003.03512",
    archivePrefix = "arXiv",
    primaryClass = "hep-th",
    reportNumber = "EPHOU-20-003, WU-HEP-20-03, KUNS-2788",
    doi = "10.1103/PhysRevD.102.045005",
    journal = "Phys. Rev. D",
    volume = "102",
    number = "4",
    pages = "045005",
    year = "2020"
}

@article{Kobayashi:2022tti,
    author = "Kobayashi, Tatsuo and Otsuka, Hajime and Sakamoto, Makoto and Takeuchi, Maki and Tatsuta, Yoshiyuki and Uchida, Hikaru",
    title = "{Index theorem on magnetized blow-up manifold of T2/ZN}",
    eprint = "2211.04595",
    archivePrefix = "arXiv",
    primaryClass = "hep-th",
    reportNumber = "EPHOU-22-019, KYUSHU-HET-249, KOBE-TH-22-05",
    doi = "10.1103/PhysRevD.107.075032",
    journal = "Phys. Rev. D",
    volume = "107",
    number = "7",
    pages = "075032",
    year = "2023"
}

@article{Ding:2026sfm,
    author = "Ding, Gui-Jun and King, Stephen F. and Qu, Bu-Yao",
    title = "{Froggatt-Nielsen mechanism with non-invertible symmetry}",
    eprint = "2610.26761",
    archivePrefix = "arXiv",
    primaryClass = "hep-ph",
    month = "10",
    year = "2026"
}

@article{Ding:2026bhp,
    author = "Ding, Grace and Chen, Yanzhu and Huang, Sheng-Jie",
    title = "{Hybrid Lattice Surgery between Smooth Boundaries}",
    eprint = "2610.27556",
    archivePrefix = "arXiv",
    primaryClass = "quant-ph",
    month = "10",
    year = "2026"
}

@article{Nakai:2025thw,
    author = "Nakai, Yuichiro and Otsuka, Hajime and Shigekami, Yoshihiro and Zhang, Zhihao",
    title = "{The minimal supersymmetric standard model with non-invertible selection rules}",
    eprint = "2512.21509",
    archivePrefix = "arXiv",
    primaryClass = "hep-ph",
    reportNumber = "KYUSHU-HET-347",
    doi = "10.1140/epjc/s10052-026-16184-z",
    journal = "Eur. Phys. J. C",
    volume = "86",
    number = "8",
    pages = "970",
    year = "2026"
}
\bibliographystyle{JHEP} 

\end{document}